\documentclass[twocolumn]{aastex631}

\hypersetup{linkcolor=red,citecolor=blue,filecolor=cyan,urlcolor=blue}
\usepackage{graphicx,color}
\usepackage{amssymb}
\usepackage{amsmath}
\usepackage{url}
\usepackage{natbib}
\usepackage{txfonts}

\newcommand{\kms}{km~s$^{-1}$}
\newcommand{\degree}{\ensuremath{^\circ}}

\newcommand{\sdo}{\textit{SDO}}

\newcommand{\hri}{HRI$_{\rm EUV}$}
\newcommand{\feix}{\ion{Fe}{9}}
\newcommand{\fex}{\ion{Fe}{10}}
\newcommand{\fe}{\ion{Fe}{9}/\hspace{-0.1cm}\ion{}{10}}
\newcommand{\ov}{\ion{O}{5}}
\newcommand{\ovi}{\ion{O}{6}}
\newcommand{\oxy}{\ion{O}{5}/\hspace{-0.1cm}\ion{}{6}}
\newcommand{\siiv}{\ion{Si}{4}}

\usepackage{url}
\usepackage{hyperref}

\begin{document}
	
	
	\title{SolO/EUI Observations of Ubiquitous Fine-scale Bright Dots in an Emerging Flux Region: Comparison with a Bifrost MHD Simulation}
	
	\author[0000-0001-7817-2978]{Sanjiv K. Tiwari}
	\affiliation{Lockheed Martin Solar and Astrophysics Laboratory, 3251 Hanover Street, Bldg. 252, Palo Alto, CA 94304, USA}
	\affiliation{Bay Area Environmental Research Institute, NASA Research Park, Moffett Field, CA 94035, USA}
	
	\author[0000-0003-0975-6659]{Viggo H. Hansteen}
	\affiliation{Lockheed Martin Solar and Astrophysics Laboratory, 3251 Hanover Street, Bldg. 252, Palo Alto, CA 94304, USA}
	\affiliation{Bay Area Environmental Research Institute, NASA Research Park, Moffett Field, CA 94035, USA}
	\affil{Rosseland Centre for Solar Physics, University of Oslo, P.O. Box 1029 Blindern, NO–0315 Oslo, Norway}
	\affil{Institute of Theoretical Astrophysics, University of Oslo, P.O. Box 1029 Blindern, NO–0315 Oslo, Norway}
	
	\author[0000-0002-8370-952X]{Bart De Pontieu}
	\affiliation{Lockheed Martin Solar and Astrophysics Laboratory, 3251 Hanover Street, Bldg. 252, Palo Alto, CA 94304, USA}
	\affil{Rosseland Centre for Solar Physics, University of Oslo, P.O. Box 1029 Blindern, NO–0315 Oslo, Norway}
	\affil{Institute of Theoretical Astrophysics, University of Oslo, P.O. Box 1029 Blindern, NO–0315 Oslo, Norway}
	
	\author[0000-0001-7620-362X]{Navdeep K. Panesar}
	\affiliation{Lockheed Martin Solar and Astrophysics Laboratory, 3251 Hanover Street, Bldg. 252, Palo Alto, CA 94304, USA}
	\affiliation{Bay Area Environmental Research Institute, NASA Research Park, Moffett Field, CA 94035, USA}

    \author[0000-0003-4052-9462]{David Berghmans}
	\affil{Solar-Terrestrial Centre of Excellence – SIDC, Royal Observatory of Belgium, Ringlaan -3- Av. Circulaire, 1180 Brussels, Belgium}

	\begin{abstract}
		
	We report on the presence of numerous tiny bright dots in and around an emerging flux region (an X-ray/coronal bright point) observed with SolO's EUI/\hri\ in 174 \AA. 
	These dots are roundish, have a diameter of 675$\pm$300 km, a lifetime of 50$\pm$35 seconds, and an intensity enhancement of 30\% $\pm$10\% above their immediate surroundings. About half of the dots remain isolated during their evolution and move randomly and slowly ($<$10 \kms). The other half show extensions, appearing as a small loop or surge/jet, with intensity propagations below 30\,\kms.
	Many of the bigger and brighter \hri\ dots are discernible in SDO/AIA 171 \AA\ channel, have significant emissivity in the temperature range of 1--2 MK, and are often located at polarity inversion lines observed in HMI LOS magnetograms. Although not as pervasive as in observations, a Bifrost MHD simulation of an emerging flux region does show dots in synthetic \fe\ images. These dots in the simulation show distinct Doppler signatures -- blueshifts and redshifts coexist, or a redshift of the order of 10 \kms\ is followed by a blueshift of similar or higher magnitude. The synthetic images of \oxy\ and \siiv\ lines, which represent transition region radiation, also show the dots that are observed in \fe\ images, often expanded in size, or extended as a loop, and always with stronger Doppler velocities (up to 100 \kms) than that in \fe\ lines. Our observation and simulation results, together with the field geometry of dots in the simulation, suggest that most dots in emerging flux regions form in the lower solar atmosphere (at $\approx$1 Mm) by magnetic reconnection between emerging and pre-existing/emerged magnetic field. Some dots might be manifestations of magneto-acoustic shocks through the line formation region of \fe\ emission.

	
	\end{abstract}
	

	\section{Introduction} \label{sec:intro}
The energy and mass loading of the outer solar atmosphere and the evolution of the magnetic field in emerging or ephemeral magnetic flux regions remain a mystery in solar physics. It is thought that much of the dynamics in emerging flux regions is powered by magnetic reconnection \cite[e.g.,][]{cheu14,van15,moor22}. The observations of the solar X-ray and Extreme Ultraviolet (EUV) corona have revealed heating events in the form of solar explosions of varying magnitudes \citep[e.g.,][]{sves76,huds91,masu94,moor01,asch02,flet11,benz17}. Thus, the energy release events occur from large-scale solar X-ray flares to small-scale EUV dot-like brightenings. 
New data from Extreme Ultraviolet Imager \citep[EUI;][]{roch20} onboard the mission Solar Orbiter \citep[SolO;][]{mull20} show a plethora of small bright dots that may be signatures of new field expanding into the upper chromosphere/lower corona with resulting magnetic reconnection and heating.	

The emerging ephemeral regions appear in the solar corona as an X-ray bright point, also known as coronal bright point (CBP) \citep{vaia73,golu74,golu77}. These are small bipolar regions of $\sim$40", live less  than 24 hr, and at a given time have an absolute magnetic flux of 10$^{20}$ Mx or less \citep{harv73_ER,hage01,kont20}. Some CBPs might form by magnetic flux convergence and cancellation \citep{prie94,long99}. The CBPs are well studied objects -- see \cite{madj19} for a detailed review. With the availability of high spatial- and temporal-resolution data, here we are able to investigate `dot-like' substructures inside a CBP.


The presence of numerous fine-scale bright dot-like structures, often referred to as bright grains, in the quiet Sun has been reported in the past in chromospheric and transition region (TR) lines \citep[e.g.,][]{mart15}. Dots are found to be present in network regions, plage areas, and active regions \citep{mart15,skog16,brya16,depo17}. These chromospheric/TR dots, observed by IRIS \citep{depo14IRIS}, are mostly roundish, live 2 to 5 minutes, move with a speed of 30 \kms, and have a size of 400--2100 km \citep{skog16}. The majority of bright grains in weak field areas such as in the quiet Sun or coronal holes were proposed to be a result of chromospheric shocks impacting the transition region \citep{mart15,skog16}. These shocks are driven from the photospheric convection and have been shown to be associated with dynamic fibrils \citep{skog16}, commonly observed in H$\alpha$ \citep{depo07_DFs}. 


Bright dots have also been observed in plage and sunspots in coronal, EUV, wavelengths  \citep{regn14,tian14,alp16,deng16,sama17}. Sunspot penumbral bright dots were proposed to be caused by magnetic reconnection between more inclined and more vertical penumbral field \citep{alp16}. Some of these could be linked to penumbral jets and/or Ellerman bombs in sunspot penumbra  \citep{tiw16,roup21}. In plage/moss regions EUV dots were proposed to be nanoflare events \citep{regn14} [see also \cite{test13,test14,test20a,wine13} and \cite{poli18}], in which magnetic reconnection in coronal loops (at apex, or near the chromospheric/TR footpoints) can appear as small localised bursts, rapidly converting magnetic energy into thermal energy \citep{parker88,prie00,asch04}. Some of these could also be a TR density and temperature enhancement due to the impact of strong downflows along the coronal loops rooted therein \citep{tian14,klei14}.

Some dot-like fine-scale explosive events having $\sim$1 min lifetime, a diameter (FWHM of the intensity profile across dots) of 800 km, and intensity enhancements of $>$100\%, were recently reported to be present in the core of an active region \citep{tiw19} observed by Hi-C 2.1 \citep{rach19}. \cite{tiw19} found these dots to be located at polarity inversion lines (PILs). Thus, they proposed those dots to be formed by magnetic reconnection accompanied by magnetic flux cancellation and/or emergence. They also noted the presence of dot-like structures at the base of surges/jets and proposed that some dots could be a part of other, extended, explosive events such as tiny loops or surges/jets, reported therein.    
	
Here we present fine-scale dot-like transient brightening events at the location of an emerging magnetic flux region observed by the telescope EUV High Resolution Imager, \hri, of EUI onboard SolO. Some of these dots could be considered as the smallest EUV brightenings, `campfires', recently reported by \cite{berg21,pane21}, but earlier by e.g., \cite{falc98}, in the  quiet solar corona. We use the commissioning phase data for this study and carefully select 170 dots in the emerging flux region and characterize them by estimating their sizes, lifetimes and intensity enhancements with respect to their immediate surroundings. We find that dots are present everywhere in the field-of-view of EUI's \hri\ 174 \AA\ observations, but they are in an appreciably higher density near stronger field regions, particularly in the emerging flux region that we investigate here. For a better understanding and interpretations of dots we also synthesize and use images in \fe, \oxy, and \siiv\ emissions from a Bifrost magnetohydrodynamic (MHD) simulation.

	\begin{figure*}[h]
		\centering
		\includegraphics[trim=0cm 1.6cm 0.81cm 0.82cm,clip,width=\linewidth]{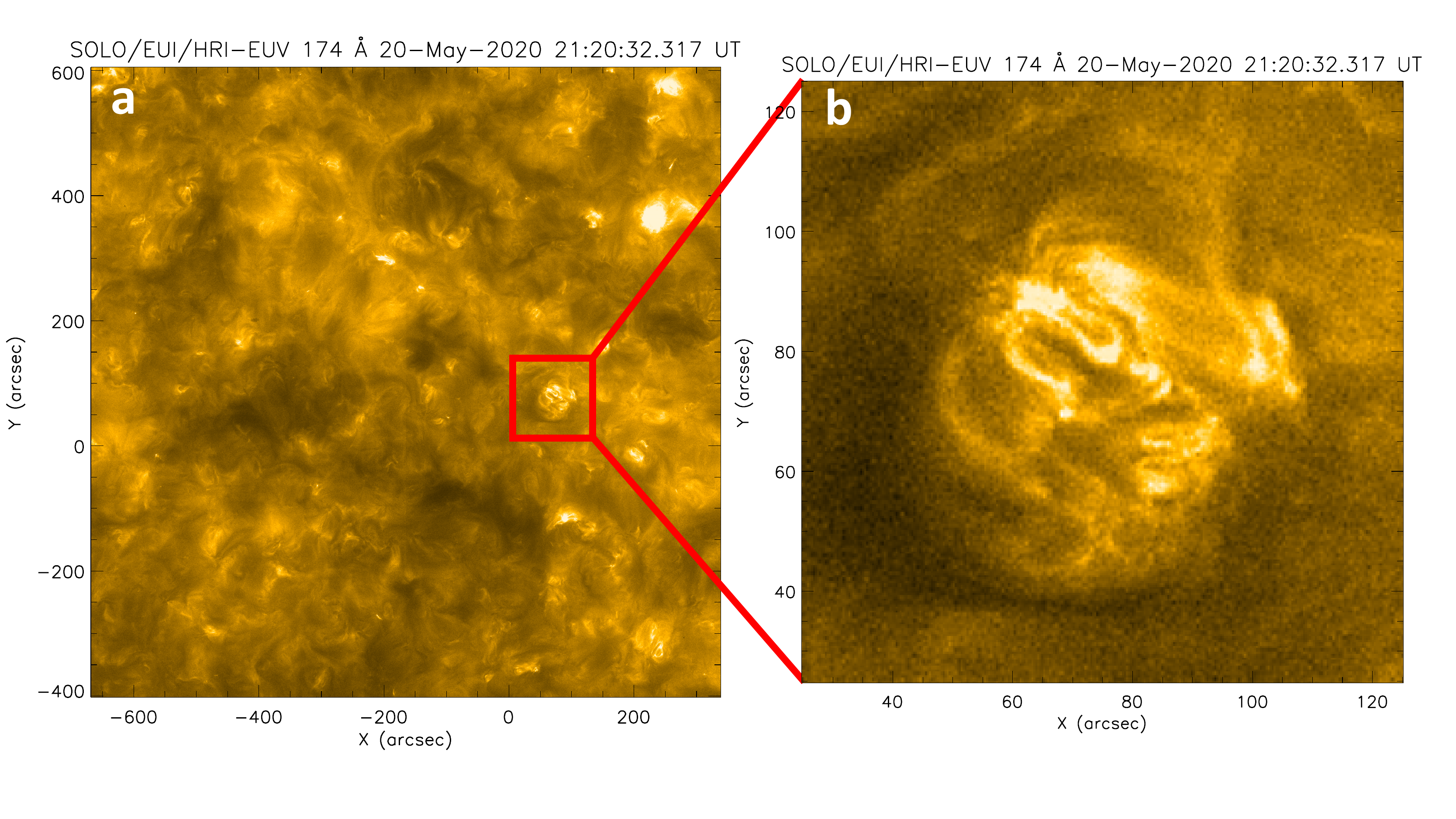}
		\includegraphics[trim=0cm 6.65cm 7.7cm 0cm,clip,width=\linewidth]{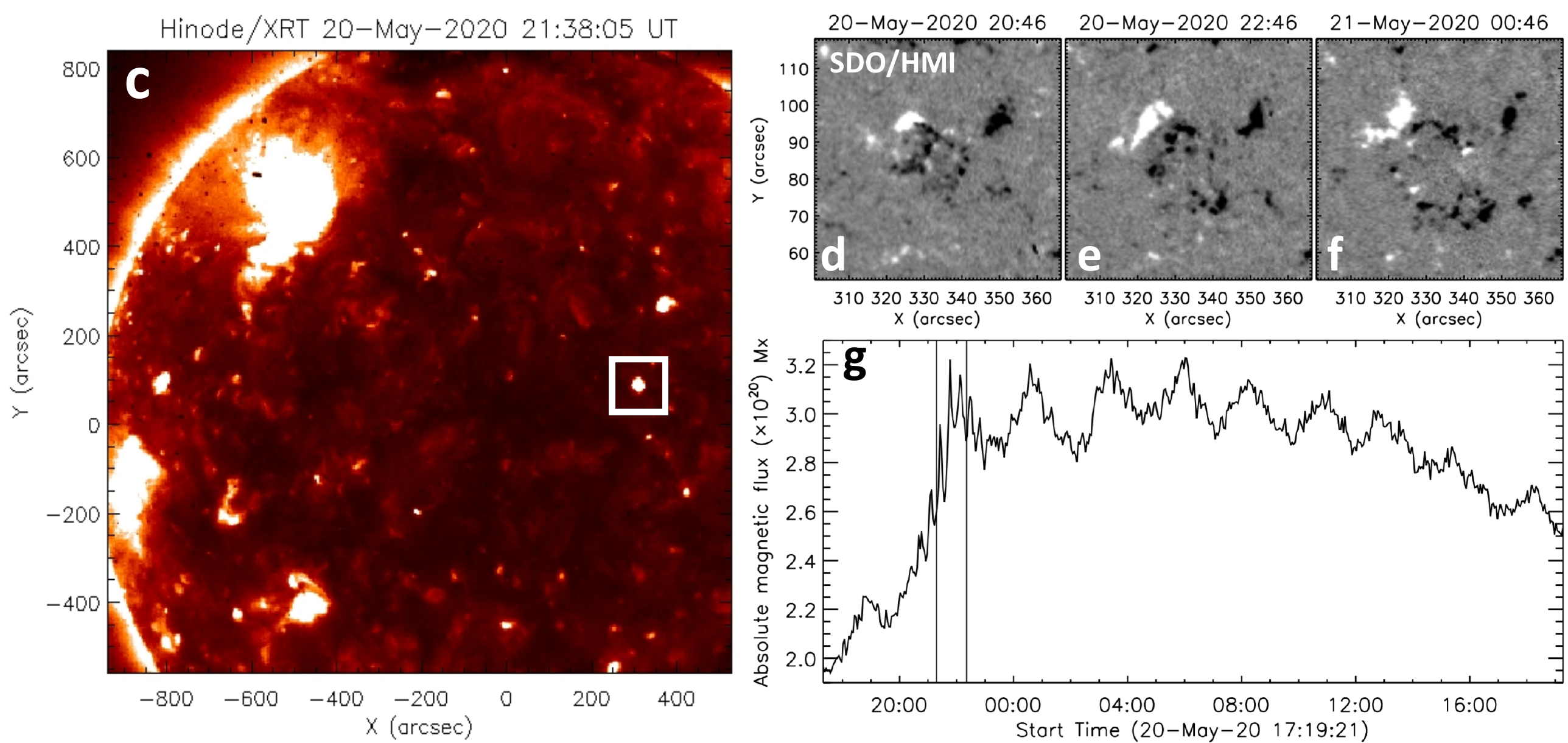}
		\caption{A context image (panel a) of the full field-of-view (FOV) observed by \hri\ on 20-May-2020 at 21:20:32 UT. An emerging flux region is outlined by a red box, an enlarged version of which is displayed in panel b and used for analysing fine-scale dots. One arcsec of the \hri\ data on 20-May-2020 covers 442 km on the Sun. Panel c shows an Hinode/XRT image in which the emerging region, outlined by a white box, appears as an X-ray/coronal bright point. Panels d--f display three maps of SDO/HMI LOS magnetic field of the emerging flux region, the magnetic flux of which over 24 h is plotted in panel g. The two vertical lines in g outline the duration of \hri\ observations. In all panels in this figure, as well in all other images in the paper, solar north is up, and solar west is to the right. } 
		\label{f1}
	\end{figure*}

	\section{Data, Methods, and Modelling}\label{sec:data}
	
	We analyse data of a small magnetic flux emergence region that was covered in the quiet Sun coronal observations of \hri\ on May 20, 2020 (Figure \ref{f1}). As evident from the Hinode/X-Ray Telescope \citep{golu07} image in Figure \ref{f1}c, the emerging flux region is a classical X-ray/coronal bright point.
Line-of-sight (LOS) magnetograms (Figure \ref{f1}d--f) and flux evolution plot (Figure \ref{f1}g) provide further evidence of this region to be a CBP with short-closed loops therein.
	We use calibrated L2 EUI data\footnote{\url{https://doi.org/10.24414/wvj6-nm32}} for our study. The L2 data product is the calibrated data, suitable for scientific analysis. 
	The \hri\ wavelength passband ranges from 171  to 178 \AA, and is centered on 174 \AA. Thus, the EUI 174 \AA\ channel detects the characteristic emissions of the \feix\ and \fex\ lines from the coronal plasma at about 1 MK. The EUI 174 \AA\ passband also includes \oxy\ lines, and thus detects some TR emission, presumably -- see more discussion on this later. The plate scale of EUI data used in the present analysis is 0.492\arcsec. The SolO/EUI was situated at 0.609 AU from the Sun on May 20, 2020, thus one \hri\ pixel corresponds to about 217 km on the Sun, with a resultant two-pixel Nyquist spatial resolution of 434 km.
	
   The telescope \hri\ obtained 174 \AA\ images between 21:20:12 and 22:17:02 UT, cycling through a two-minute program. During this period \hri\ took 5 images at a 10 s cadence plus a 6th image 70 s later. Thus, Solar Orbiter provided  the \hri\ images  with aforementioned  temporal cadence for about  57  minutes.	These observations were taken as a part of a technical compression test of \hri. Therefore, these images have variable settings. Nonetheless, during the 57-minute observations 60 images were well exposed, un-binned and compressed at high quality levels. Our analysed data frames are from within these 60 high quality images.  
   	We have used a cross-correlation technique to co-align \hri\ images with each other. However, because of the variable compression the co-alignment using the cross-correlation method has not been straightforward and thus the co-alignment cannot be considered perfect.
	
	Because of the usual enhanced brightness in coronal loops in the emerging flux region, dots mostly appear faint. Therefore, we created unsharp masked images, from \hri\ 174 \AA\ images, to enhance the visibility of dots, see e.g., Figure \ref{f2}. For this purpose we subtracted smoothed frames (by a factor of 5 pixels) from the original data. We have prepared movies of \hri\ 174 \AA\ images and unsharp masked images, available as online animation (``movie1.mp4"). In the movie we have kept all image frames available during the 57-minute observations, for reference, and have not removed binned and/or bad frames.   
	
We also use EUV data obtained with Atmospheric Imaging Assembly \citep[AIA;][]{leme12} on-board Solar Dynamics Observatory \citep[SDO;][]{pesn12}, and LOS magnetograms from the Helioseismic and Magnetic Imager \citep[HMI;][]{scho12, sche12}, also onboard SDO. A similar unsharp masking, using 5$\times$5 AIA pixels, is applied to AIA 171 \AA\ images. Note that Solar Orbiter was at 0.609 AU from the Sun on 20-May-2020, therefore, the events would appear 3.22 minutes earlier in SolO/EUI images than in the SDO/AIA images. The angle between SolO and Sun-Earth line on 20th May 2020 was 16.4\degree. 
All our generated maps (\hri, AIA and HMI) were processed and de-rotated using SolarSoft routines \citep{free98}. 
The reference frame for de-rotation is the central image in our SDO data, i.e., at 20-May-2020 21:44:51 UT. A roll angle correction of 6\degree\ is made to match that with SolO/EUI. The solar (X,Y) SDO coordinates at the reference time are as follows: xrange=[300,365], yrange=[53,118].  The magnetograms are within 30\degree\ from the disk center, and therefore a projection effect correction is not essential \citep[see, e.g.,][]{falc16}.
A movie (``movie2.mp4") containing AIA 171 \AA\ images and its unsharp masked images, together with HMI LOS magnetograms, corresponding to Figure \ref{f4_sdo}, is available online.

\subsection{Selection criterion of dots} 
\begin{figure*}
	\centering
	\includegraphics[trim=3.75cm 3.45cm 2.58cm 1.9cm,clip,width=\linewidth]{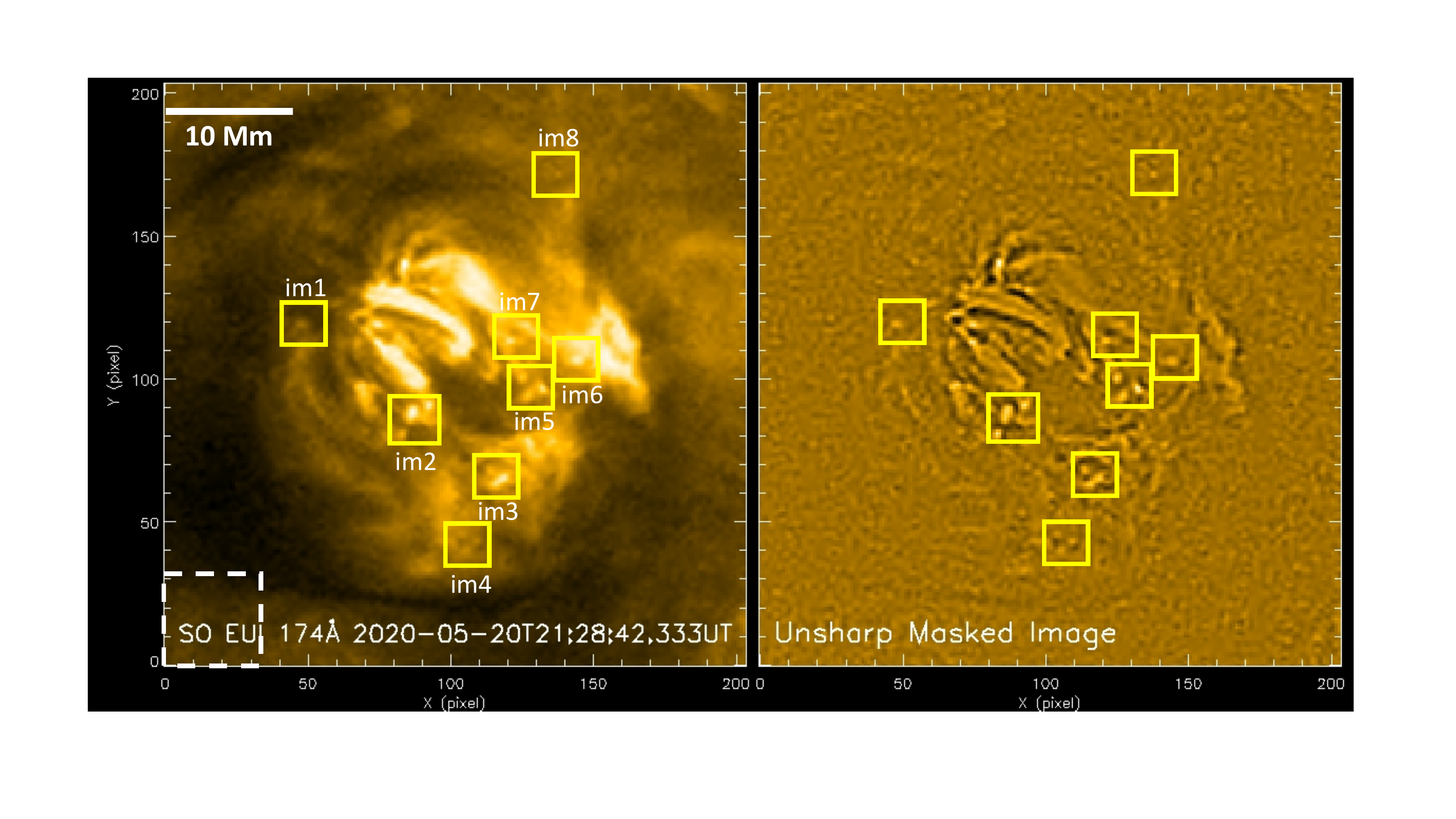}
	\includegraphics[trim=0.73cm 0.5cm 2.4cm 3.5cm,clip,width=\linewidth]{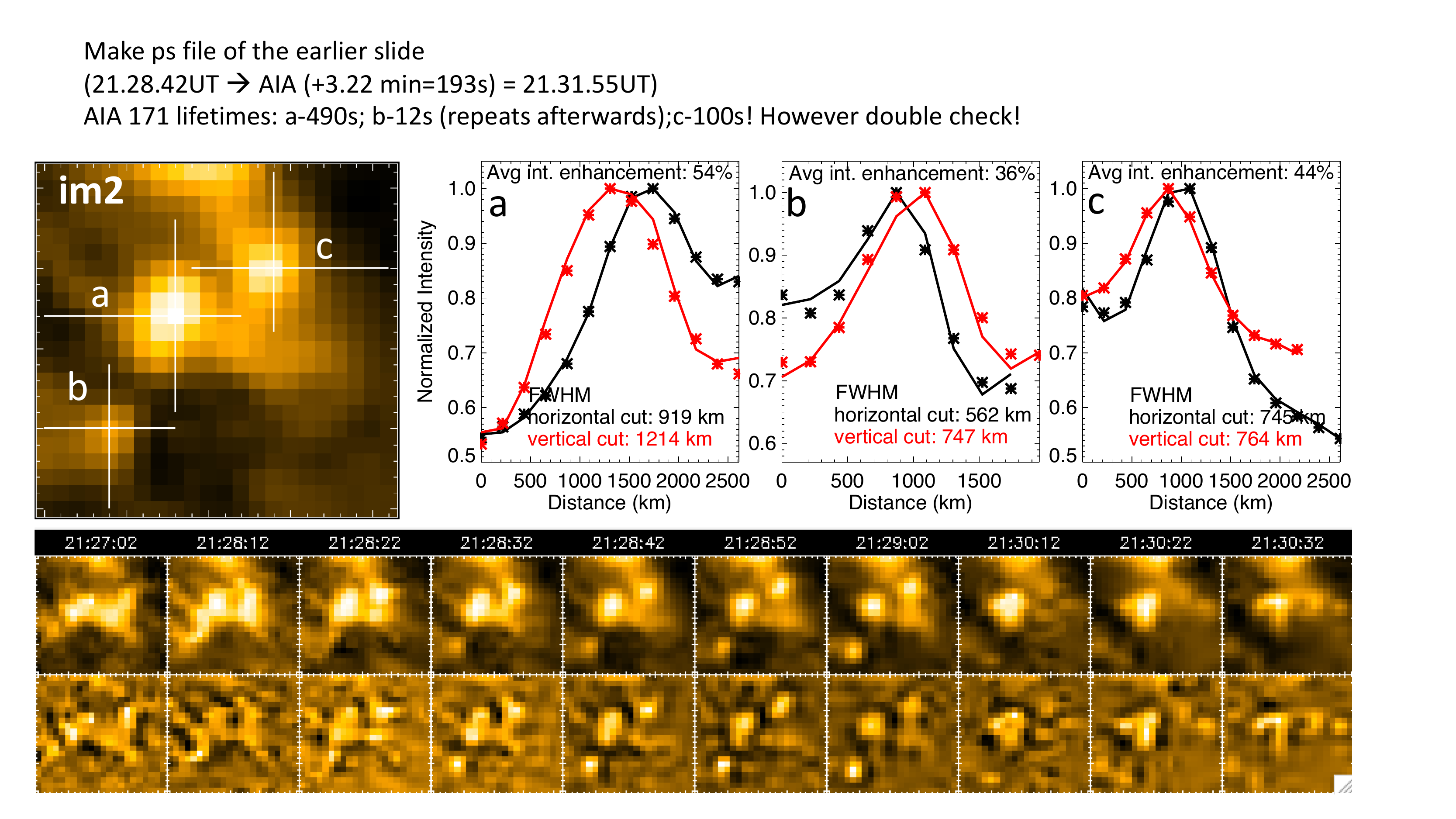}
	\caption{Examples of fine-scale dots in EUI/\hri\ observations. Top row: the left panel is an \hri\ 174 \AA\ image  of the same FOV as the Figure \ref{f1}b, and the right panel is an unsharp masked image of it; different boxes outline the regions of selected dots in this image frame; a white dashed box on the bottom left outlines the region that is used for noise estimation; a white horizontal bar on the top left scales 10 Mm distance, for reference. 
		Middle row: on the left is a zoomed in view of one of the boxed regions, named `im2' of the same time as that in the top row. This region has three dots, named a, b, and c. The sizes along the horizontal (in black) and vertical cuts (in red) of these dots, together with their Gaussian fits (asterisks) are shown in the right three panels.  The profile plots are on the original \hri\ images. The brightness enhancement of each dot with respect to their immediate surroundings is also printed. Bottom row: Time series images, together with the corresponding unsharp masked images, of the region outlined by box `im2' are shown to follow the evolutions and lifetimes of the three dots a, b and c. Note that the images are not isochronal. This is because this \hri\ data set was taken as a part of a compression test, as described in Section \ref{sec:data}. An animation of the uppermost row is available online (``movie1.mp4"). The movie has same FOV but no annotations, and it runs (in SolO time) from 21:20 to 22:17 UT. }
	\label{f2}
\end{figure*}

We employ two criteria for selecting dots: (i) We select a quieter region in the surroundings of the emerging region (e.g., outlined by a dashed white box in Figure \ref{f2}) and estimate the mean value of the intensity and its standard deviation inside the box, which is considered the 1$\sigma$ noise level.  We definite dots that have an intensity enhancement above 2$\sigma$ level from their surrounding. For example, if 1$\sigma$ of the quiet region is 8\% of the mean for an image frame, all selected dots in that image frame should have intensity enhancements above 16\% of their immediate surroundings. The second criterion is: (ii) the dot should be visible in at least two consecutive image frames.
Thus, if a dot is at or above 2$\sigma$ level intensity from their immediate surrounding, and is visible in two or more consecutive image frames, the dot has been considered for analysis. This ensures that the selected dots are not noise.

However, we have considered a few exceptions. We found a few striking dots to be above the 3$\sigma$ intensity level of their surroundings but visible in only one image frame, thus having a lifetime of less than 10 s. We have included those\footnote{There is no way to rule out the possibility of these dots being the effect of cosmic rays. Nonetheless, these dots present less than 5\% of our sample and do not affect the main results.}. 

Similarly, a few dots whose intensity enhancement is slightly below or close to the threshold 2$\sigma$ value, but are visible in two or more consecutive image frames, are also included. Together, these dots represent less than 10\% of our sample. The reason for including these is that these dots are most likely real, but don't qualify the criterion either because of a lower cadence of the data than their lifetimes, or due to them not being strikingly bright with respect to their surroundings, which are also bright. Another factor that we considered for keeping these dots in our sample is that, as mentioned above, there is a 70 s gap after five consecutive image frames, which complicates finding the true lifetimes of dots.   

The selected dots are isolated enough from other dots and bright structures in their surroundings to characterize them more accurately. 
Dots are mostly roundish (unlike penumbral dots [see, e.g., \citet{tian14} and \citet{alp16}], which are often extended along sunspot penumbral filaments [\citet{tiw13}]). Thus, we take horizontal and vertical cuts for measurement of the size of dots. Even when dots are extended in a direction there is no preferred direction. On the one hand, this suggests that the dots in the emerging flux region might be different from the `elongated' dots in the sunspot penumbra. On the other hand, this also implies that any dot extensions would not be caused by camera artefacts. We therefore keep it simple by measuring distances along two cuts (in horizontal and vertical directions) and then by averaging them to extract the diameter of each dot.
We selected and characterized each of the dots in \hri\ manually because the dots are usually dim with respect to their surroundings, and the \hri\ pointing is not always stable during this
commissioning phase, so that automatic selection of these fine-scale dots would fail and/or give erroneous results.

\subsection{Bifrost MHD Model}\label{bifrost}

We use a Bifrost MHD simulation of an emerging flux region. This is a new simulation in that we have modelled the quiet Sun network/an emerging magnetic flux region by injecting a horizontal flux sheet of time-varying strength at the bottom boundary of a model that spans a domain of $72\times 72\times 61$~Mm$^3$ on a grid of $[720,720,1115]$ grid points using the Bifrost code \citep{gudi11}. The model reaches from 8.5~Mm below the photosphere and extends into the corona, up to 52.5 Mm above the photosphere. The Bifrost model includes optically thick radiative transfer, including scattering in the chromosphere \citep[see, e.g.,][]{skar00,haye10}, and optically or effectively thin radiative transfer in the middle chromosphere to corona following the recipes of \cite{carl12}. The equation of state, including partial ionisation of the atmospheric plasma, is treated in local thermodynamic equilibrium (LTE) through a look-up table constructed using solar abundances. Furthermore, thermal conduction along the magnetic field, especially relevant for the corona, is included in the energy equation.

The model is initialised with a horizontal field of 100~Gauss up to the photosphere, with a nearly 0~Gauss field in the corona. This is evolved with an initial field injection of $B_y=200$~Gauss at the bottom boundary for 95 minutes. Convective dynamics lead to a tangled field and a hot corona $> 1$~MK after an hour of solar time. At 95~minutes the field strength of the flux sheet entering the bottom boundary was increased to $B_y=1000$~Gauss for 70 minutes followed by another increase to 2000~Gauss for the next 150 minutes. After this strong field injection is completed, the field strength injected was reduced to  $B_y=300$~Gauss, at which point it remains constant. After the first hour or so, most coronal transients have dissipated and the photospheric field closely resembles the measured photospheric field in observations (Hansteen et al. 2022, in prep). The first signs of flux emergence occur at (roughly) 3 hours, but this may be convectively processed ambient field breaking through the photosphere. Field stored just below or rising to the photosphere will break through the surface and enter the upper atmosphere once the gradient of the sub-photospheric field strength becomes sufficiently large \citep{arch04}. Stronger flux emergence occurs at later stages in the simulation -- this phase of the simulation is featured in this paper, which is suitable for comparison with the emerging flux region of \hri\ observations used in this study.

For line synthesis, we calculate their emission by integrating the contribution function  $\phi(u,T) n_{\rm e} n_{\rm H} G(T,n_{\rm e})$ along the line of sight, where $\phi$ is the emission profile, $n_{\rm e}$ and $n_{\rm H}$ are the electron and hydrogen number densities respectively, and $G(T,n_{\rm e})$ is a function describing the ionisation and excitation state of the emitting ion taken from CHIANTI \citep{dere97}. The latter assumes the ionisation and excitation equilibrium. This integration is performed using CUDA, i.e., a parallel computing platform developed by NVIDIA for computing on graphical processing units (GPU), which accelerates the integrations drastically. Since the wavelengths of the iron lines is short there is the possibility of absorption from neutral gas. We include this effect in our calculation by multiplying the contribution function along the line of sight with $\exp(-\tau)$ where $\tau$ is the combined opacity of hydrogen and helium, as well as from singly ionised helium \citep[see][for details]{depo09}.

\section{Results} 
	
\subsection{Dots in EUI Observations}	\label{results_euidots}
	
In Figure \ref{f2} we display an example \hri\ image frame and its unsharp masked image. Many of the fine-scale dots are discernible. Some of the outstanding dots are outlined by yellow boxes. The dots are particularly clearly visible in the unsharp masked image. Many more fine-scale dots are outlined in Appendix A (Figure \ref{f12}) in two additional image frames from the \hri\ movie of the emerging flux region (movie1.mp4).
We manually selected 170 dots from different image frames of the ``movie1.mp4" and characterized them by calculating their sizes, lifetimes, and intensity enhancements with respect to their immediate surroundings. Most of the fine-scale dots are roundish in the image frame when they are selected for analysis. While half of the dots remain roundish during their lifetimes, the other half of dots extend, sometimes explosively, to become a loop or a jet/surge-like event (see examples in Figure \ref{f2}). Sometimes a dot splits into two or more dots, and occasionally two or more dots merge to become a single dot or a slightly extended structure.

\begin{figure*}
	\centering
	\includegraphics[trim=0cm 0cm 0cm 0cm,clip,width=\linewidth]{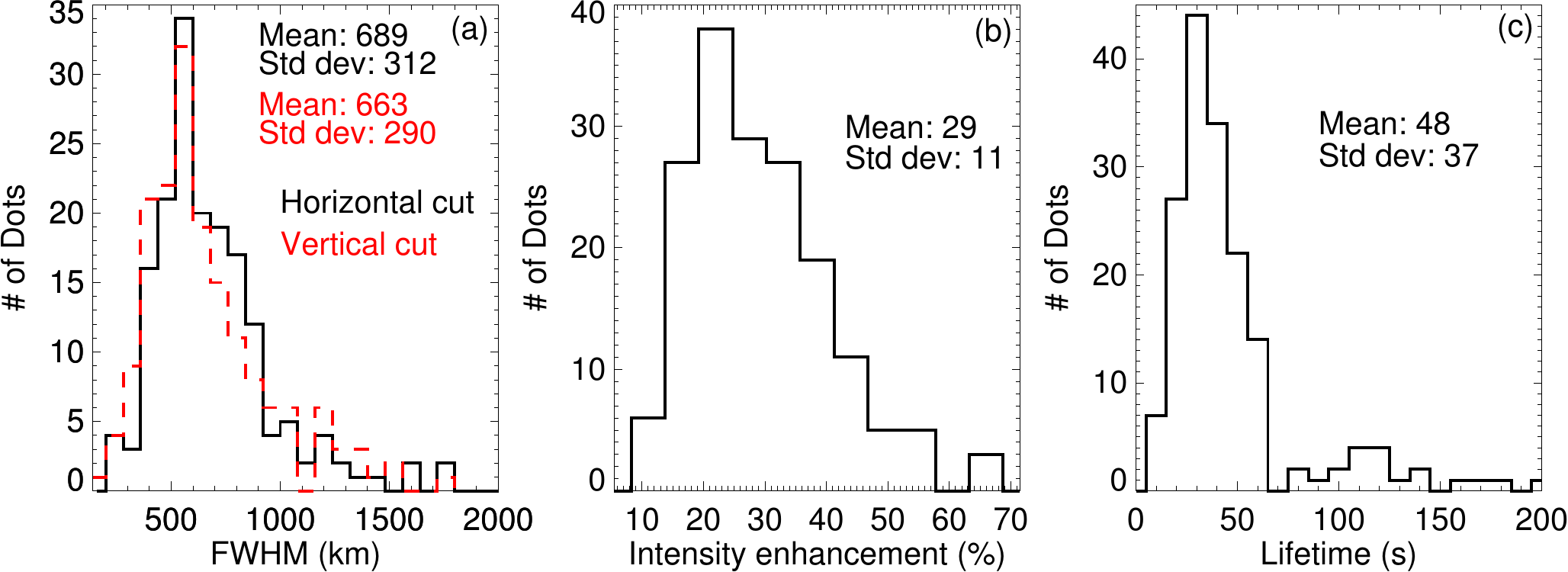}
	\caption{Histograms of the sizes, intensity enhancements, and lifetimes of 170 dots observed with \hri. Panel (a) shows histograms of the horizontal and vertical sizes of dots. Panel (b) shows histogram of intensity enhancements with respect to the immediate surroundings of dots. Panel (c) displays the histogram of lifetimes of dots. } 
	\label{f3_hist}
\end{figure*}

The estimation of intensity enhancements of dots with respect to their background is done by averaging the minimum intensity values at each end of the two intensity profiles, separately, that are used for assessing the horizontal and vertical sizes of dots. Then we evaluate what percentage of this averaged value, for each profile, is the peak intensity value. An average of these two percentage numbers (from the two intensity profiles) is the percentage brightness enhancement of a dot with respect to its immediate surroundings. The following general caveat should be kept in mind when interpreting and/or comparing the percentage intensity enhancement of different dots from their immediate surroundings (as measured here, and generally done in the studies cited in this paper): the percentage intensity increase depends on the background, which can be different from instrument to instrument (affected by e.g., stray-light, telescope's point spread function/PSF), from wavelength to wavelength, and from environment (active region, ephemeral region, quiet sun) to environment.

In Figure \ref{f2}, we also display three dots `a', `b, and `c' from inside the box `im2', more closely. Dot `a' is the biggest (diameter $\sim$1000 km) and brightest (54\% brighter than its immediate surroundings) of the three, and extends towards the south becoming a jet-like activity. Dot `b' is the smallest (diameter $\sim$650 km) and dimmest (36\% brighter than its immediate surroundings) of the three, and remains mostly isolated until its disappearance.  Dot `c' appears to originate from a longer bright loop-like structure in its south (see the image at 21:28:12 UT).

Intensity enhancements might be considered underestimated to some extent due to the fact that we have taken averages of the intensities of four locations/pixels at minimum intensities for the two cuts (horizontal and vertical) and have not taken averages of many more pixels in the surroundings that could have lower values than the four points along the two cuts, particularly  because most the emerging region is bright, in general.
	
For the lifetime of each dot we follow the dot manually in a zoomed-in region and visually find out the time between when the dot looked as a dot and when it disappears, or becomes another feature such as a loop or a jet/surge.
Examples of the estimation of the lifetimes of three dots are described in the last row of Figure \ref{f2}. 

\subsubsection{Statistical properties of HRI$_{\rm EUV}$ dots}

We performed a statistical quantitative analysis of the physical properties of dots observed by \hri, and created their histograms. 
The histograms of the sizes, lifetimes and intensity enhancements of 170 dots are displayed in Figure \ref{f3_hist}. 
Most dots have a diameter of about 800 km or less. Thus, most dots are fairly small in size. But some dots can have a diameter of as large as 2000 km, or more. The averages of horizontal and vertical sizes of the 170 dots come out to be 689$\pm$312 km and 663$\pm$290 km, respectively. Although some dots appear extended in one direction (during their measurements -- when they are most circular), evidently the dot's horizontal and vertical sizes are not significantly different. Thus, the average diameter of the dots in our sample is 676$\pm$301 km.    

Most dots have an intensity enhancement of 20--40\% to their immediate surroundings, with an average of 29$\pm$11\%. Because our dots are dim, many of these dots are not so clearly visible in direct \hri\ 174 \AA\ images, and become clearer in the unsharp masked images. The intensity enhancement of dots is rather low as compared to the previously reported numbers of brightness enhancements for EUV dots in the literature. For example, these numbers are fairly low as compared to that of the dots reported in the core of an  active region using Hi-C 2.1 data, that have an intensity enhancements of $\ge$ 100\%\footnote{Please note that the brightness enhancements for Dot 1 and Dot 2 in Figure 3 of \cite{tiw19} (and any corresponding text) should read as 120\% and 435\%, respectively, which were inadvertently given as of 60\% and 80\%.}. 
But again, remember the caveat mentioned in Section \ref{results_euidots} on how the estimation of percentage intensity enhancement can be affected by different magnetic backgrounds, wavelengths, and instruments.

\begin{figure*}
	\centering
	\includegraphics[trim=1.5cm 0cm 1.46cm 0cm,clip,width=\linewidth]{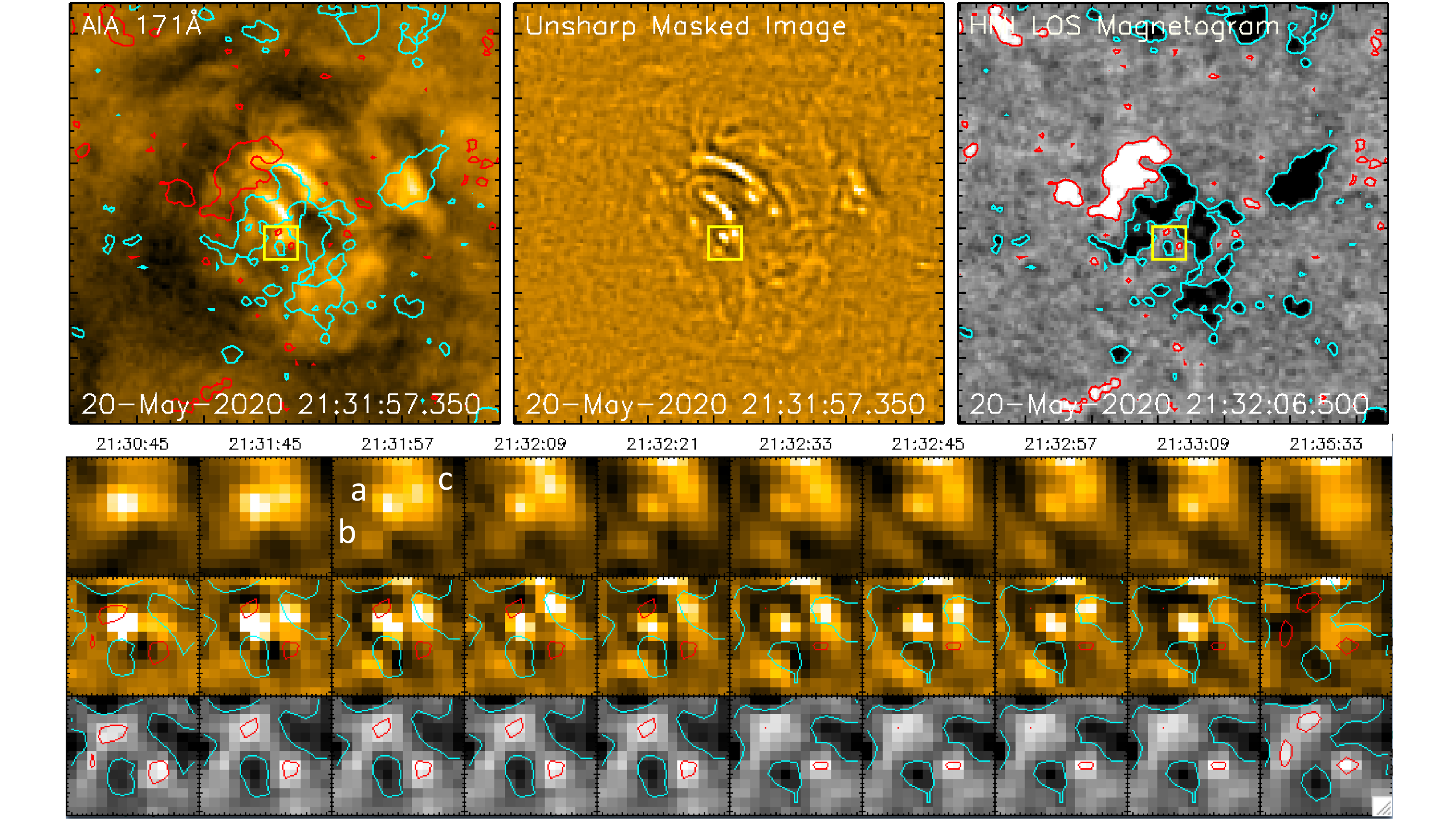}
	\caption{\sdo\ observations of the same emerging flux region for the corresponding time of the \hri\ image displayed in Figure \ref{f2}. The top panel shows, from left to right, AIA 171 \AA\ image, its unsharp masked image, and HMI LOS magnetogram, respectively. The (X,Y) coordinates are same as in Figure \ref{f1}d--f.
		A yellow box outlines the same region as im2 in Figure \ref{f2}. The AIA 171 \AA\ image and LOS magnetogram have HMI LOS magnetic field contours (of the same time) of level $\pm$20 G overlaid -- red is for positive magnetic polarity and cyan is for negative magnetic polarity. The LOS magnetogram is saturated at $\pm$40 G. The bottom panel shows the evolution of the same three dots a, b, and c, as in Figure \ref{f2}, but here as seen from SDO/AIA. The last row of the bottom panel contains corresponding LOS magnetic field evolution together with its contours from SDO/HMI.
			An animation ``movie2.mp4" is available online. Its annotations and FOV are same as in the upper panel of the figure and it runs from 21:20 to 22:20 UT at a 12 s cadence.  }
	\label{f4_sdo}
\end{figure*}

The dots are fairly short-lived. The lifetime of most dots is below a minute, but some can be longer-lived, up to three minutes or more. Most dots in the previous literature have longer lifetimes, on average, than that of the dots reported here. The average lifetime of our \hri\ dots is 48$\pm$37 s, which can be considered as the upper limit for average duration of dots. This is because, as described in Section \ref{sec:data}, the \hri\ images are not isochronal, so that if a dot's lifetime is longer than 70 s we can not be sure if it is the same event or there are two subsequent events.
 
We also measured the plane-of-sky (PoS) speeds of a few dots that show considerable proper motions. We also calculated the speeds of brightness propagation along dot's extension by creating time-distance maps in the way described later in Section \ref{proper_motion}; see Figure \ref{hori_speed}. Most dots show a random and slow proper motions of about 2 -- 10 \kms. The brightness propagation along dot's extension can have a speed of up to 30 \kms. Because most dots show a little proper motion, and there are some image alignment issues in the commissioning phase data, it is often difficult to track and quantify the speeds of fine-scale structures such as dots. Note that a proper motion of 2 \kms\ is comparable to the displacements caused by uncertainty in the image alignment.

\begin{figure*}
	\centering
	\includegraphics[trim=0cm 0cm 0cm 0cm,clip,width=\linewidth]{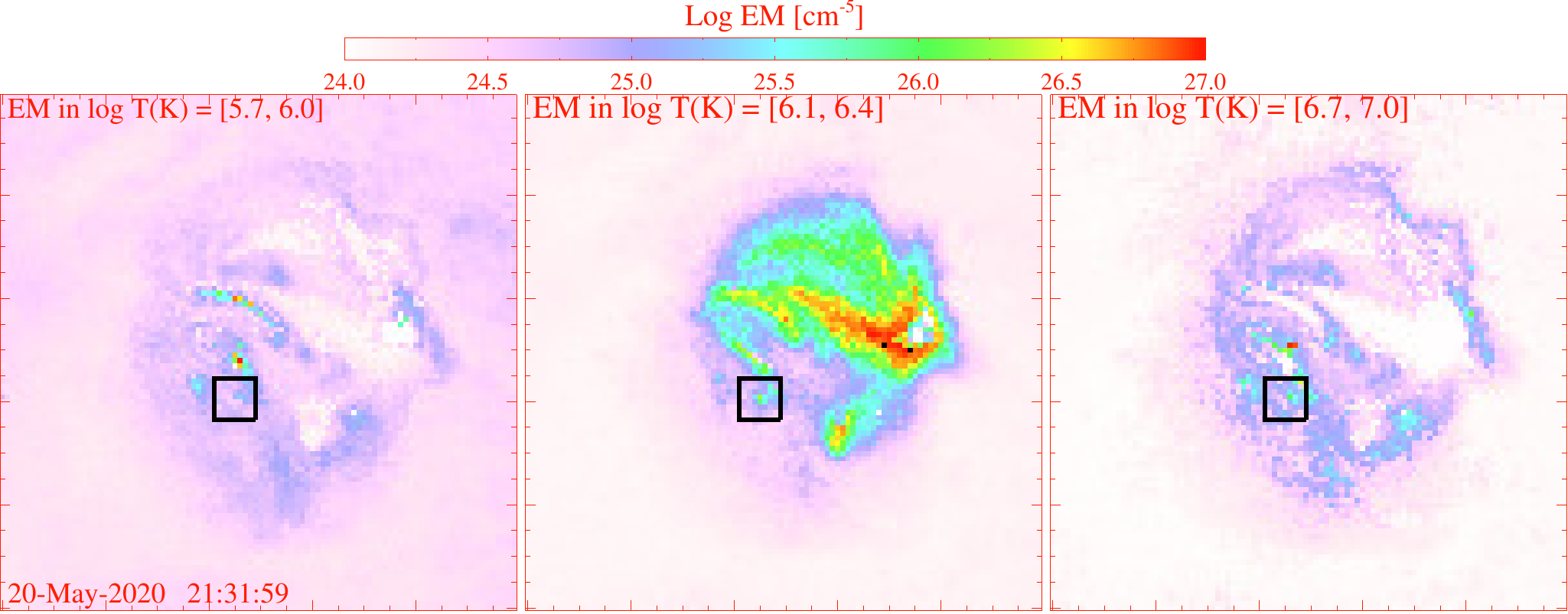}
	\caption{The EM maps, of the same FOV as AIA 171 \AA\ image in the upper panel of Figure \ref{f4_sdo}, in three log temperature (log$_{10}$ T) bins. The black box outlines the same region as im2 in Figure \ref{f2}, and also outlined by a yellow box in \sdo\ images in Figure \ref{f4_sdo}, covering dots a, b, and c. Two (a and c) of the three dots apparently have significant EM in the temperature range of 1--2.5 MK. The third one (dot b) does not have significant emission in any of the temperature ranges shown.} 
	\label{dem}
\end{figure*}

\subsection{Dots in SDO Observations}	
	
We manually co-aligned \sdo\ (AIA and HMI) data with EUI \hri\ data to allow tracking of fine-scale EUI dots in AIA 171 \AA\ images. We found that many of the \hri\ dots are discernible in \sdo/AIA 171 \AA\ images. Because the angle between LOS directions of SolO and SDO during our observations was only 16.4\degree, there was no foreshortening correction made. 
These dots obviously exceed 1 AIA pixel, similar to campfires \citep[e.g.,][]{rutt20,berg21,pane21}.

We followed the evolution of magnetic flux, underlying the dots, in HMI LOS magnetograms. Figure \ref{f4_sdo} shows an AIA 171 \AA\ image, and its unsharp masked image, nearly at the same time as the \hri\ image in Figure \ref{f2}. The difference in the noted times in the two images (EUI image is 195 s earlier than AIA) is due to the difference in the distance of the two spacecrafts from the Sun. The HMI LOS magnetogram of the nearest time to the AIA 171 \AA\  image is also shown, with magnetic contours over plotted on it. The contours are also over-plotted on the AIA 171 \AA\ image to allow easy tracking of magnetic polarity distribution underlying the dots. 
Because these dots are often dim in AIA 171 \AA\ and are not as outstanding as they are in \hri\ 174 \AA\ images, these were not reported and particularly explored in an emerging flux regions in the past in AIA observations. 

The three dots (`a', `b', and `c') in AIA 171 \AA\ image, outlined by a yellow box in Figure \ref{f4_sdo}, are the same as those shown inside box ``im2" in Figure \ref{f2} observed with \hri.
Although the dots are not so obvious in AIA 171 \AA\ images, these, and many of the other bigger and brighter  \hri\ dots, become discernible in the unsharp masked images of SDO/AIA 171 \AA\ channel. In Figure \ref{aia_channel}, we display images, of the same FOV as in Figure \ref{f4_sdo}, in several AIA channels. We find faint signatures of dots `a', `b', and `c' in AIA 304, 193, and 131 \AA\ images, similar to that seen in AIA 171 \AA\ image. A distinct signature of these dots cannot be seen in AIA 1600 and 1700 \AA\ images. These together further suggest that many, if not all, dots are TR feet of coronal loops.

The lower row of Figure \ref{f4_sdo} shows that mixed-polarity magnetic flux is present near the dots a and c -- emerging minority-polarity (positive) magnetic flux eventually cancels with pre-existing majority-polarity (negative) magnetic flux. The movie ``movie2.mp4" with AIA and HMI data show that the biggest dots, such as dot `a' in Figure \ref{f4_sdo}, are often seated at or near sharp neutral lines, and apparent emergence and cancellation of minority-polarity magnetic flux is seen. This behaviour of magnetic field distribution is similar to that seen for Hi-C 2.1 dots in the arch filament system in the core of an active region by \cite{tiw19}, who also found flux cancellation rates of the order of 10$^{16}$ -- 10$^{17}$ Mx s$^{-1}$ for some of their fine-scale explosive events. Note that there is no signature of mixed-polarity magnetic flux near dot `b'. This could be due to the opposite-polarity flux elements being beyond the detection limit of HMI. Or, the dot `b' might not form the same way as dots `a' and `c'. 

Similar to that what is seen in the \hri\ images during evolution of dots, we also note that the dots in AIA images show extensions as a loop, or a jet, mostly during their later phase, but occasionally dots also form at the end of an explosion (from a more extended, jet-like or loop-like, structure).

\subsubsection{Differential emission measure}		
		
To determine the emission of dots in different temperature bins we performed DEM analysis by using six AIA channels (171, 211, 335, 193, 94, and 131 \AA) following the method described in \cite{cheu15dem}. 
In Figure \ref{dem}, emissivity for three temperature bins are displayed at approximately the same time as for the images in Figures \ref{f2} and \ref{f4_sdo}. Two (a and c) out of the three dots outlined by the box in Figure \ref{f4_sdo} (which are the same ones from \hri\ inside box `im2') show emissivity in the temperature bin of log$_{10}$T = [6.1,6.4]. The third, the smallest dot in the south (named as `b'), does not show any emission in any of the T bins. Thus, this dot could be at a cooler chromospheric/TR temperature. However the two larger dots are evidently heated to a coronal temperature of a MK or more.

\begin{figure*} [t]
	\centering
	\includegraphics[trim=0cm 0.05cm 0cm 0cm,clip,width=0.96\linewidth]{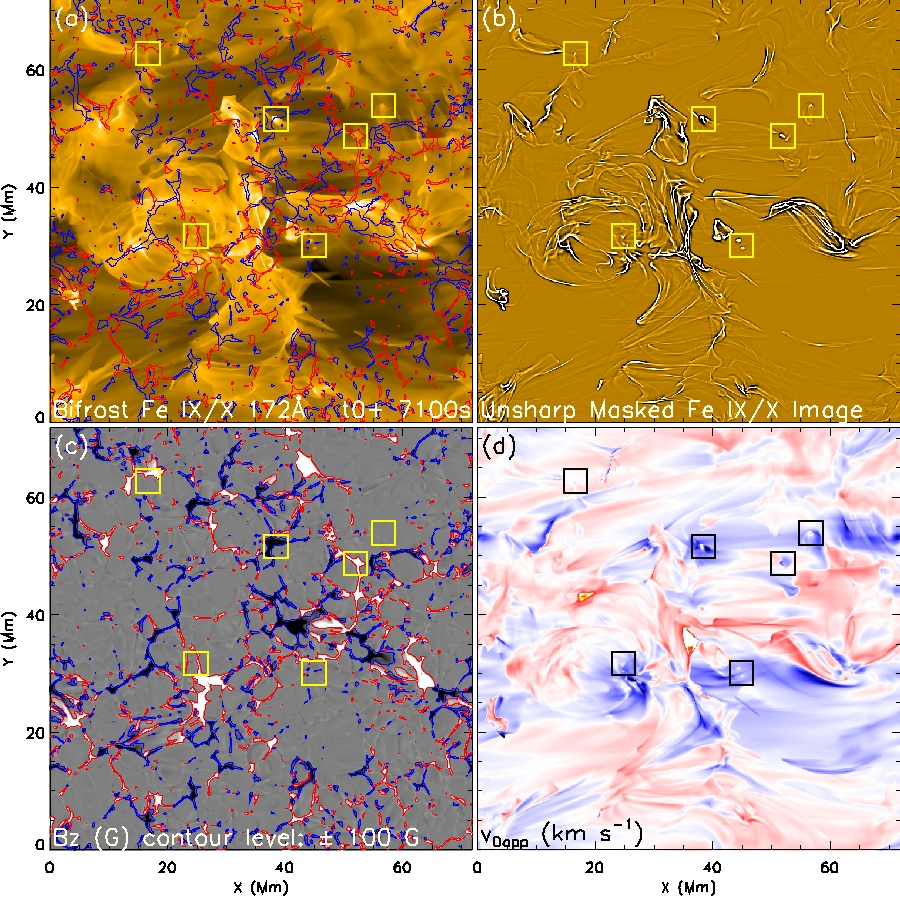}
	\caption{Images of three physical parameters from the Bifrost MHD simulation of an emerging bipolar region: (a) Synthesized \fe\ emission, (b) unsharp masked image of panel (a), (c) Surface Bz map, saturated at $\pm$500 G, and (d) \fe\ Dopplershift, v$_{Dopp}$ (representing the observable), with the upper and lower values saturated at $\pm$50 \kms. The contours (of $\pm$100 G) of the magnetogram are over-plotted on (a) and (c). A few dots are outlined by yellow boxes (black in Dopplergrams), for examples. 
	}
	\label{f6_sim}
\end{figure*}

We performed this analysis on several AIA image frames corresponding to \hri\ observations and found similar results. The biggest and brightest dots show up in the temperature range of 1--3 MK, or sometimes even at higher temperatures. However, the dimmer and cooler ones are not visible in the 1 MK or higher temperature bins. A caveat is that because of the limited temperature sensitivity of the low/TR temperatures in the AIA channels the TR contributions to the AIA passbands are relatively poorly constrained. Further, dots are so short-lived that a statistical equilibrium assumed in the DEM analysis might not be valid.

\begin{figure*} 
	\centering
	\includegraphics[trim=0cm 0cm 0cm 0cm,clip,width=\linewidth]{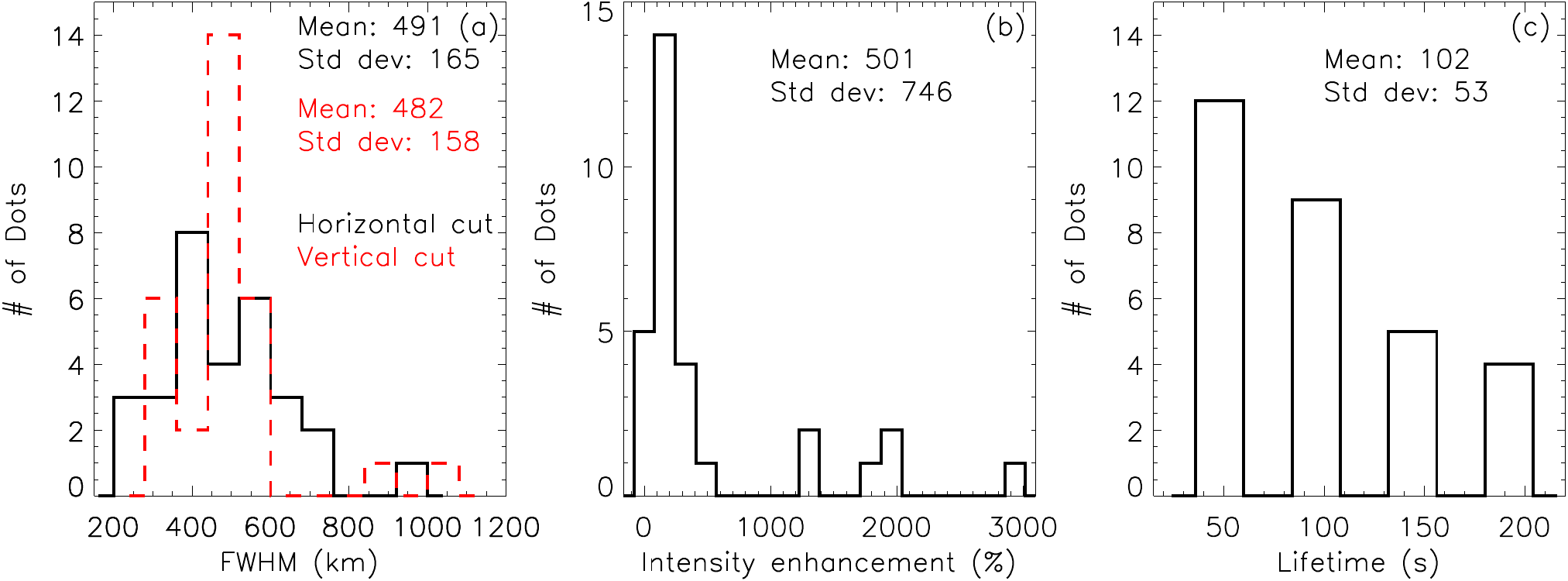}
	\caption{Histograms of the sizes, intensity enhancements, and lifetimes of 30 dots from our Bifrost MHD simulation. Panel (a) shows histograms of the horizontal and vertical sizes of dots. Panel (b) shows histogram of intensity enhancements with respect to the immediate surroundings of dots. Panel (c) displays the histogram of lifetimes of dots. } 
	\label{sim_hist}
\end{figure*}

\subsection{Dots in Bifrost MHD simulation -- synthesized \fe\ emissions}	
	
We use synthesized \feix\ and \fex\ emissions from the Bifrost simulation, described in Section \ref{bifrost}, and calculated the line intensities, Doppler speeds v$_{Dopp}$, and vertical component of the magnetic field Bz. We mainly chose \feix\ and \fex\ lines because these are the pivotal lines in the \hri\ 174 \AA\ passband. The \feix\ 171 \AA\ and \fex\ 174 \AA\ are optically thin lines.
To keep the similarity with the EUI observations we averaged synthetic \feix\ and \fex\ emissions to make a single \fe\ map that is then used for the analysis. We also averaged Dopplergrams of the two lines for consistency. We used the simulation frames from 420 ($\sim$5.8 h after the start) to 567 ($\sim$7.8 h after the start) at a 50 s cadence. We use these frames because these appear to match best with the magnetic field evolution in the early phase of the flux emergence that is captured by the \hri\ observations used here. The pixel size of the simulation is 100 km.  

In Figure \ref{f6_sim}, an example frame of the simulation containing an \fe\ emission image, its unsharp masked image, Bz, and  v$_{Dopp}$ map, are shown, all in top-down vertical simulation viewing. Several of the outstanding dots are outlined by yellow (black in Dopplergrams) boxes. Details of Bifrost dot Dopplergrams can be found in Section \ref{bifrost_dopp}.
We investigated 30 dots in detail and find that, similar to those in the observations, half of the dots in the simulation show extension during their evolution, to become a loop or jet-like activity. 
We estimated the sizes, lifetimes, and intensity enhancements of dots from their immediate background in the same way as done for the dots in \hri\ observations, and plotted their histograms in Figure \ref{sim_hist}.	For a reasonable comparison, we have estimated lifetimes, sizes, and brightness levels on the synthetic \fe\ data after smearing them to the EUI resolution. Nonetheless, as we did not use the telescope's PSF to smear the synthetic images, the comparison is still not completely fair.

We find the sizes of dots in the simulation (487 km, on the average) to be about 28\% smaller, on the whole, from that found for dots in \hri\ observations (see Figures \ref{f3_hist}a and \ref{sim_hist}a). Partly this could be due to not using the telescope's PSF to smear the synthetic images to \hri\ spatial resolution. 
The lifetimes of dots in the simulation are on the higher end from the observations -- most dots in the simulation live 50 -- 100 s.  Some appear only in a frame and have a lifetime of less than 50 s. Please note that the lower limit of the lifetime of dots is the cadence of our simulation, i.e., 50 s. 
The intensity enhancements of the dots from their surroundings vary significantly. For most dots the brightness level is several 100\% from their immediate surroundings, with an overall average of 500\%. This is much higher than the intensity enhancements of the dots in the emerging flux region of \hri, but is of the order of the brightest Hi-C 2.1 dots (found in the core of an active region; \citealt{tiw19}). 
In Figures \ref{f7_simdots1} and \ref{f8_simdots2} we show the evolution of different physical parameters of five example dots. Dot 1 is an isolated dot that remains roundish throughout. Dot 2 is also an isolated dot but extends briefly to become a loop-like (or jet-like when seen in \oxy\ and \siiv\ lines) structure. This dot presents most obvious mixed-polarity magnetic flux at its base. Dot 3 is mostly isolated and splits into two before disappearing. Dot 4 is initially a loop or surge like brightening, extended in the south, that contracts to become a dot. Dot 5 explodes to become a surge/jet like activity.

\subsubsection{Synthetic \oxy\ and \siiv\ Emissions}

Although we mainly focus on \fe\ intensities for majority of our analysis, we created images of synthetic \ov\ and \ovi\ emissions, as well as \siiv\ emissions, to see whether dots are conspicuous in these cooler transition region lines. The main reason for synthesizing \oxy\ lines is that the EUI \hri\ 174 \AA\ and SDO/AIA 171 \AA\ passbands do contain \ov\ 172.2 \AA\ and \ovi\ 173 \AA\ lines \citep[e.g.,][]{delz11}. We averaged \ov\ and \ovi\  emissions to create a \oxy\ map, in the same way as done for \fe\ emissions.

\begin{figure*}
	\centering
	\includegraphics[trim=5.6cm 0.001cm 6.6cm 0cm,clip,width=\linewidth]{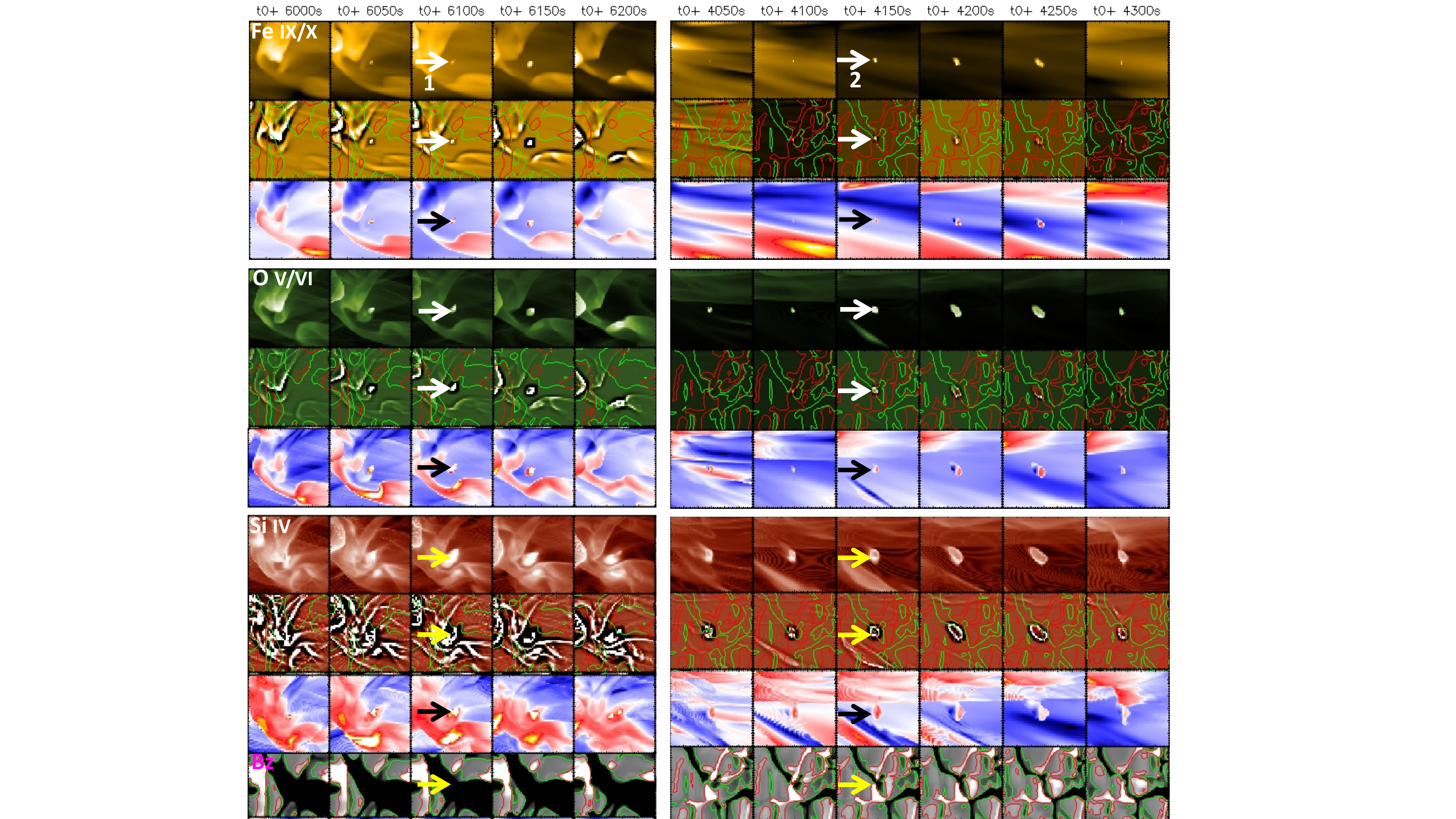}
	\caption{The evolution of two example dots, named as 1, and 2, in the synthetic \fe\ images. These dots are marked by white (or yellow and black for better visibility) arrows when they are most roundish. The top three rows represent \fe\ intensity image, its unsharp masked image, and Doppler velocity map, respectively. The same maps for \oxy\, and \siiv\ lines are  plotted in the middle and bottom three rows, respectively. There is an additional row at the bottom of the bottom row containing vertical component of the magnetic field, Bz. The Bz contours (of level $\pm$ 20 G) are over-plotted on the Bz maps and on the unsharp masked images of \fe\ emission. The FOV is $\sim5\times5$ Mm$^2$. 
	} 
	\label{f7_simdots1}
\end{figure*} 

\begin{figure*}
	\centering                    
	\includegraphics[trim=5.3cm 0.01cm 7.85cm 0cm,clip,width=\linewidth]{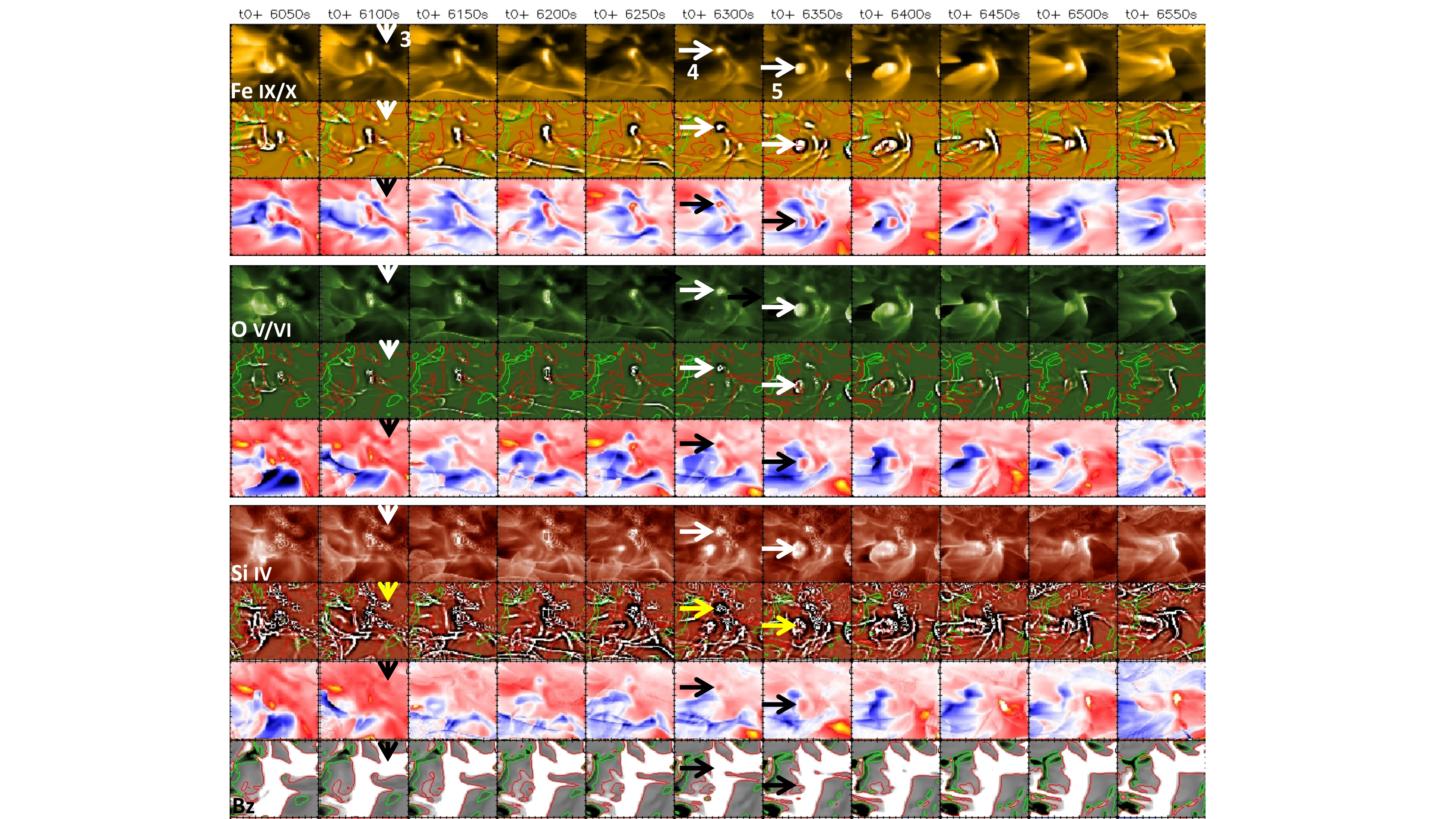}
	\caption{Same as Figure \ref{f7_simdots1}, but displaying a different region with three additional examples of dots, namely 3, 4 and 5.  } 
	\label{f8_simdots2}
\end{figure*}

\begin{figure*}
	\centering                    
	\includegraphics[trim=1.1cm 0cm 1.6cm 0cm,clip,width=\linewidth]{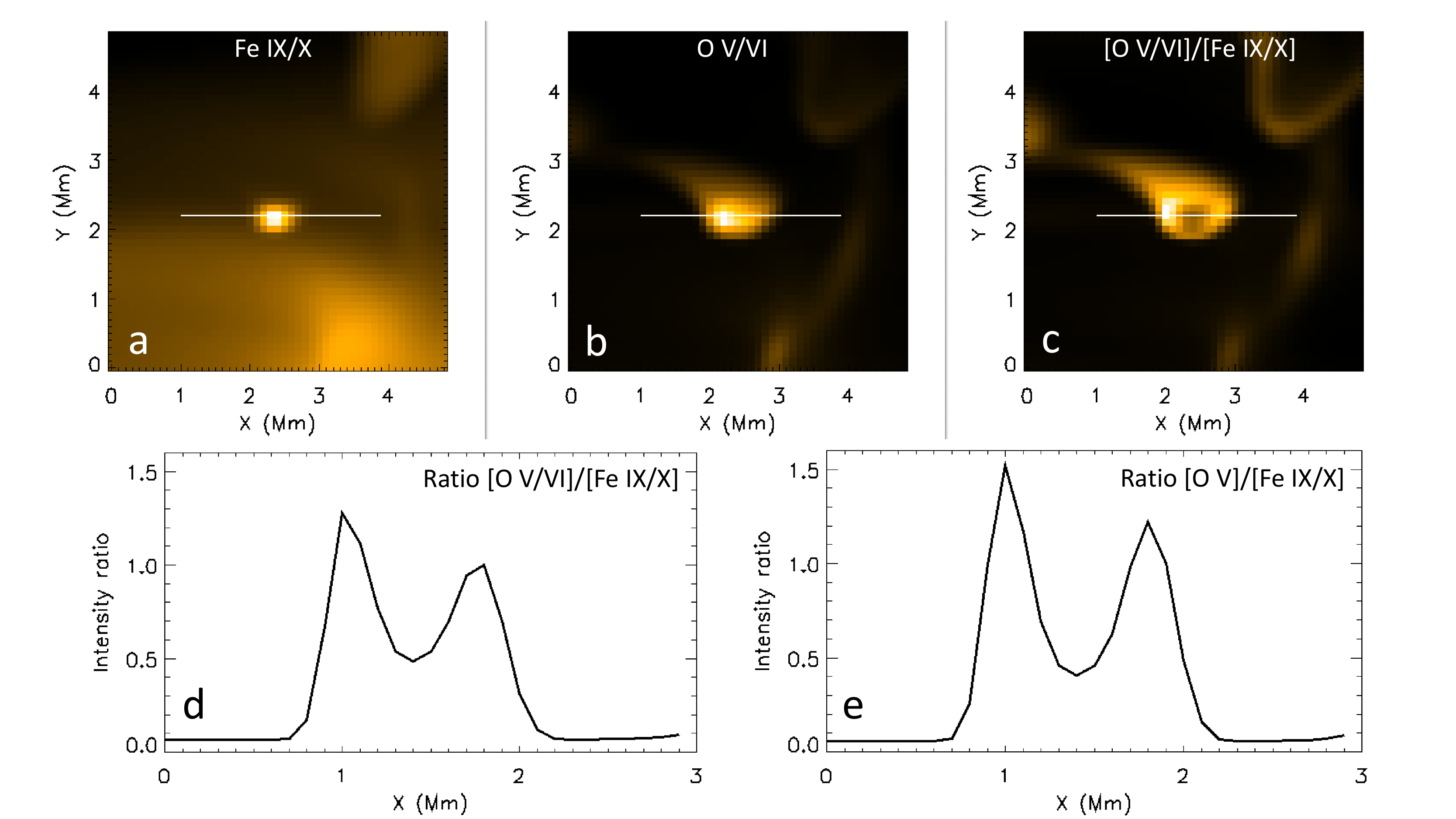}
	\caption{Line ratios of \oxy\ to \fe\ for a typical dot from the Bifrost MHD simulation. Panels a and b show the images of the dot in \fe\ and \oxy\ lines. The panel c contains the image of the ratio of \oxy\ to \fe\ lines. A horizontal line in each panel crosses the dot in the horizontal direction. Panels d and e show the intensity profiles of the ratios [\oxy]/[\fe] and [\ov]/[\fe] along the horizontal line in panel c. The intensity profiles suggest that the \oxy\ lines are equally strong to the \fe\ lines and could play a significant role in the appearance of dots observed with \hri.}
	
	\label{line_ratios}
\end{figure*}

The \ov\ and \ovi\ lines form at temperatures of 280,000 K and 320,000 K, respectively. But \ovi\ line profile has a very long tail towards higher temperatures, and it is often more coronal like in its appearance \citep{dere97}. Thus, the averaged \oxy\ images cover temperatures from about 200,000 K  to up to a million K. The \siiv\ line forms in the transition region at about 80,000 K. Thus, these lines cover the cooler to hotter transition region and lower corona. IRIS SJ 1400 \AA\ images show similar dots in an emerging flux region to that seen in \hri\ and in the synthetic \oxy\ and \siiv\ images -- see an example image from IRIS SJI in 1400 \AA\ in Appendix Figure \ref{iris_dots}. Such bright dots have also been observed in CBPs in Mg  II k and C II SJIs \citep{kays17}. Similar dots have also been observed by \cite{rutt17}
in magnetic network concentrations -- there most bright SJI 1400 \AA\ grains coincide with the magnetic concentrations seen in the far wing of H$\alpha$. A detailed analysis of the dots in this region of IRIS observations is beyond the scope of the present paper but will be presented in a follow up study. 
Figure \ref{osi_context} shows the images in \oxy\ and \siiv\ emissions corresponding to that in Figure \ref{f6_sim} for \fe\ lines. 

To find out whether dots in the simulation are really at mostly MK temperature or at TR temperature we estimated how bright the \oxy\ emission is in the dots relative to the \fe\ emission. The ratios \oxy\ to \fe\ and \ov\ to \fe\ for an example dot are shown in Figure \ref{line_ratios}. The ratio plots (Figures \ref{line_ratios}d and e) suggest that the dots have a significant contribution from the TR emission. The dot obviously also has coronal emission. This suggests that dots are multi-thermal, possibly formed at low heights. Nonetheless, ``low height" is only a
	conjecture because we assume that the TR is under the corona. In reality, TR could just come from the
outer part of a feature (like a jet) that is cooler inside,
as in \cite{hill19}.


When we compare the five example dots of \fe\ emissions from Figures \ref{f7_simdots1} and \ref{f8_simdots2} with those in \oxy\ and \siiv\ emissions, we can notice that the dots are either bigger, expanded in size (dots 2, 3 and 5) in the TR lines, or are extended in one direction along a loop or jet-like structure (dots 1 and 4). In particular when one follows their evolution, half of the dots extend as a loop or a surge/jet. This behaviour is consistent with flaring arch filaments (FAFs) reported by \cite{viss15b}. A closer look at AIA 1600 and 1700 \AA\ images do not show enhanced activity associated with dots in  Figure  \ref{aia_channel}, thus questioning these dots being related to FAFs. Obviously, this subject requires a further detailed investigation using simultaneous UV and EUV observations. 

Further, some dots can either be a part of a loop or tiny loops themselves. Most dots in \fe\ emission appear to be at the chromospheric/TR base of a loop. 
We analyse different other properties of the simulated dots in the \fe, \oxy, and \siiv\ emissions. For example, we investigated their Doppler speeds, proper motions, and magnetic field distributions together with their 3D magnetic configurations, given below, that were not possible with the available observations.

\subsubsection{Proper motion of dots and the speed of intensity propagations in \fe\ emission}	\label{proper_motion}

\begin{figure*}
	\centering
	\includegraphics[trim=0cm 7.7cm 3.5cm 0cm,clip,width=\textwidth]{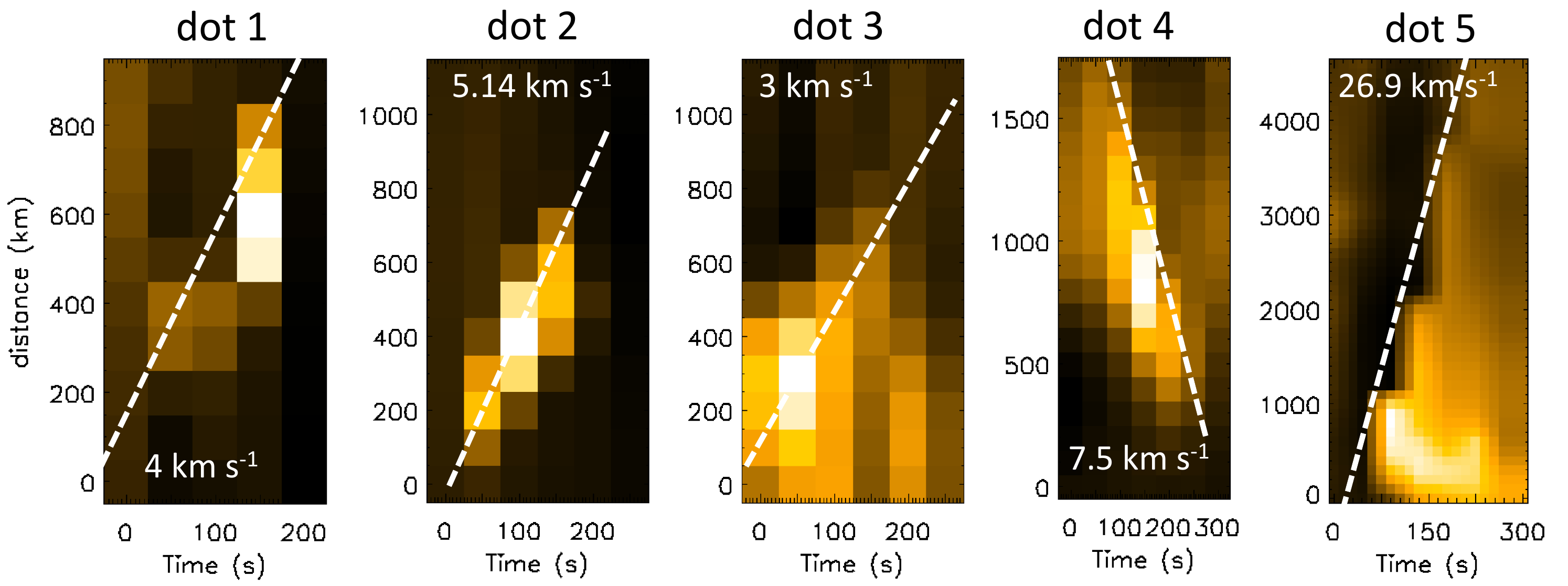}
	\caption{Time-distance maps, along the longer extensions of each of the five example dots from the simulation, to demonstrate how we estimated the brightness propagation speeds of dots. The dashed white line in each panel is to guide the eye along the intensity propagation. The estimated speeds are printed on each map. }
	\label{hori_speed}
\end{figure*}

We quantified the proper motion and/or intensity propagation of dots. For this purpose, we created time-distance maps along the longer extension of each dot during their evolution. In Figure \ref{hori_speed}, we show time-distance maps of five example dots (from Figures \ref{f7_simdots1} and  \ref{f8_simdots2}) to illustrate how the speeds were calculated.
For the dots that remain isolated and don't show any extension we only measured their proper motion, if any.
Most of the dots move very little themselves, having a speeds of $\le$ 10 \kms, but intensity propagations in them can be as fast as 30 \kms. However, the majority of dots shows a speed of $<$ 10 \kms\ in their intensity propagation as well. These speeds are consistent with those in observations (estimated for 20 randomly selected dots). These intensity propagation speeds are on the lower end of those found by \cite{mand21} for brightenings in a quiet solar coronal region, and show similarities with  loop-like and jet-like fine-scale explosive events seen by Hi-C 2.1 in an active region  \citep{tiw19}.

We note that dot 1 shows back and forth motion during its 2nd and 3rd frames (after 50th second), after showing an initial displacement in the first 50 s. We took the speed along its path of the longest extension. Similarly, dot 3 shows random motion in three directions -- first slightly towards north, then towards south and then towards west. The estimated speed in Figure \ref{hori_speed} is when the dot moves towards west because then it showed the most significant displacement. The speeds in the first two steps were each at 2 \kms. Several other dots show similar back and forth or random motion -- this could be driven by magneto-acoustic waves from the lower atmosphere. 

Dot 5 shows an explosive surge-like behaviour with the fastest brightness propagation, of the five examples, at a speed of 28 \kms. Dot 4 also shows a smooth unidirectional plasma flow. There are several of such dots that show unidirectional and several show bidirectional plasma flow (back and forth, not at the same time) and intensity propagations. This finding is similar to that found in short loops and surges of Hi-C 2.1 observations \citep{tiw19}.

\subsubsection{Bifrost dot Dopplershifts}\label{bifrost_dopp}

The dots in our simulation exhibit distinct Doppler shifts, not only in \fe\ emission but also in \oxy\ and \siiv\ emissions, see, e.g.,  qualitative pictures in Figures \ref{f7_simdots1} and \ref{f8_simdots2}. In fact, the strength of Doppler speed increases in \oxy\ and \siiv\ lines.
Some dots exhibit only redshifts (e.g., dot 1). A few dots in our sample contain only blueshifts in \fe\ images (e.g., dot 3). However, more commonly dots exhibit mixed Doppler shifts i.e., redshift and blueshift next to each other, see e.g., dots 2 and 5. Also for dot 4, the redshift is surrounded by blueshifts, but blueshifts are not as isolated as in dots 2 and 5.    
Consistent with their intensity images, Dopplergrams of dots do show an expansion of dots in \oxy\  and \siiv\ lines as compared to their appearance in \fe\ lines. The v$_{Dopp}$ maps show extended flows along the dot's longer extension appearing as a loop or a surge/jet.

\begin{figure*}
	\centering
	\includegraphics[trim=0cm 0.6cm 0cm 0.6cm,clip,width=0.75\linewidth]{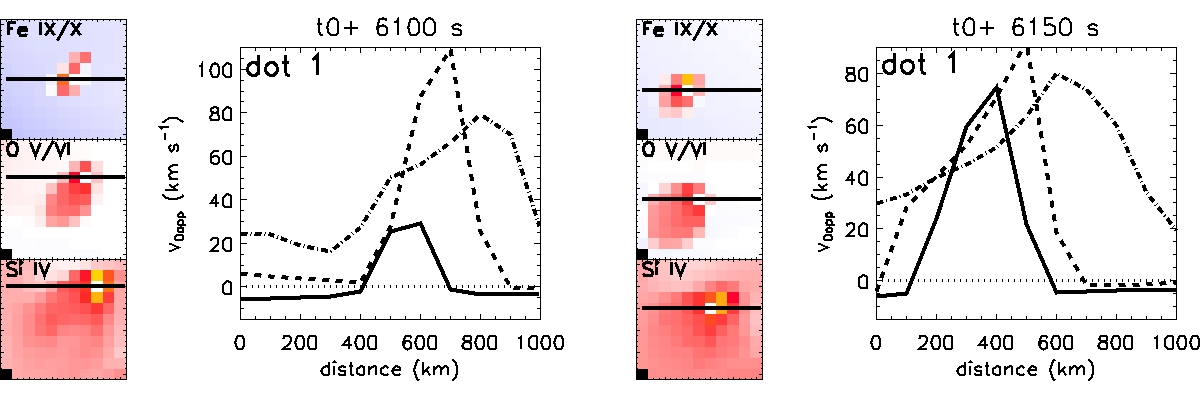}
	\includegraphics[trim=0cm 0.6cm 0cm 0.5cm,clip,width=0.75\linewidth]{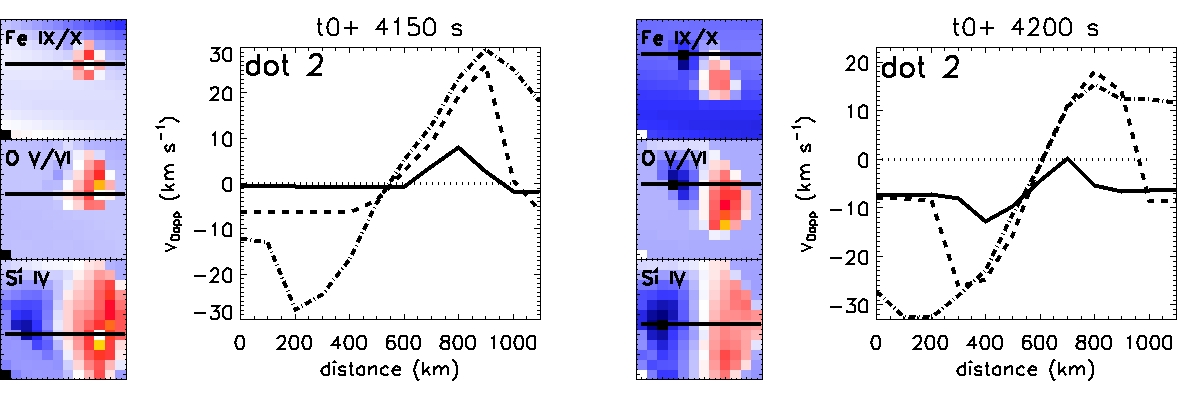}
	\includegraphics[trim=0cm 0.6cm 0cm 0.5cm,clip,width=0.75\linewidth]{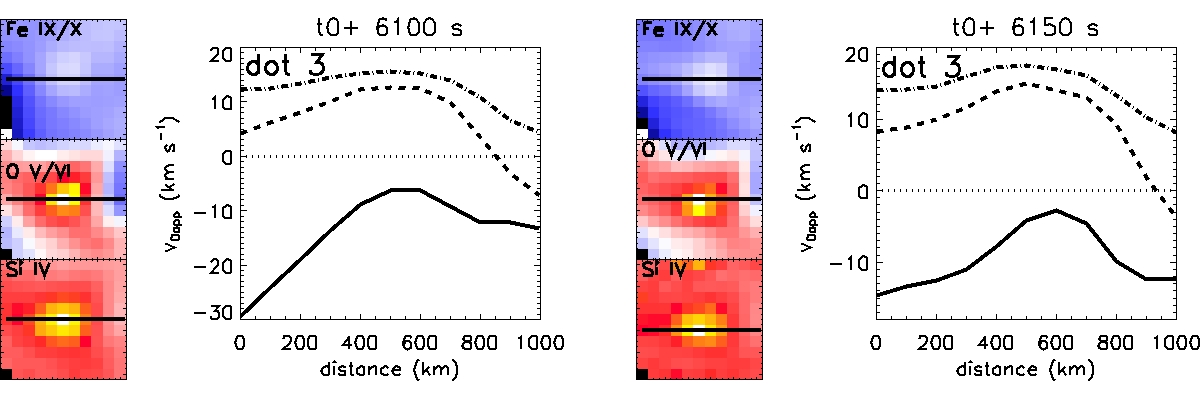}
	\includegraphics[trim=0cm 0.6cm 0cm 0.5cm,clip,width=0.75\linewidth]{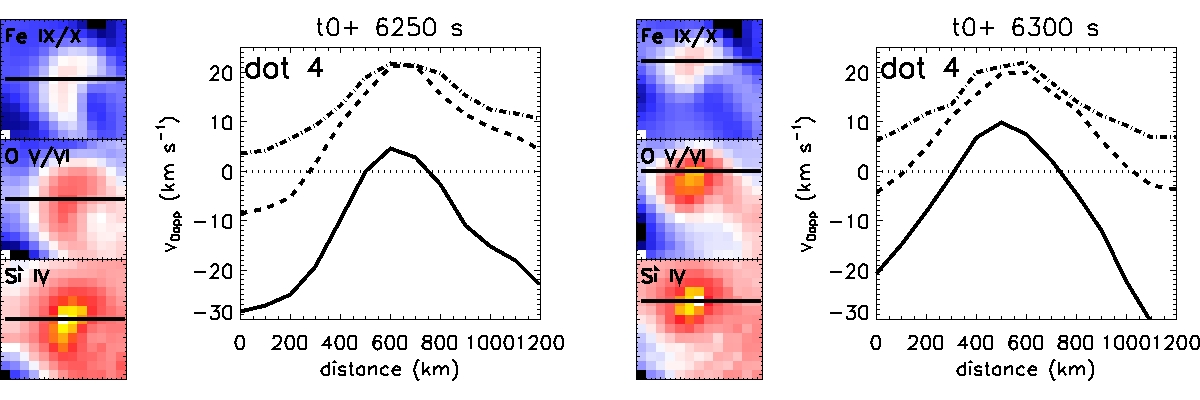}
	\includegraphics[trim=0cm 0.7cm 0cm 0.5cm,clip,width=0.75\linewidth]{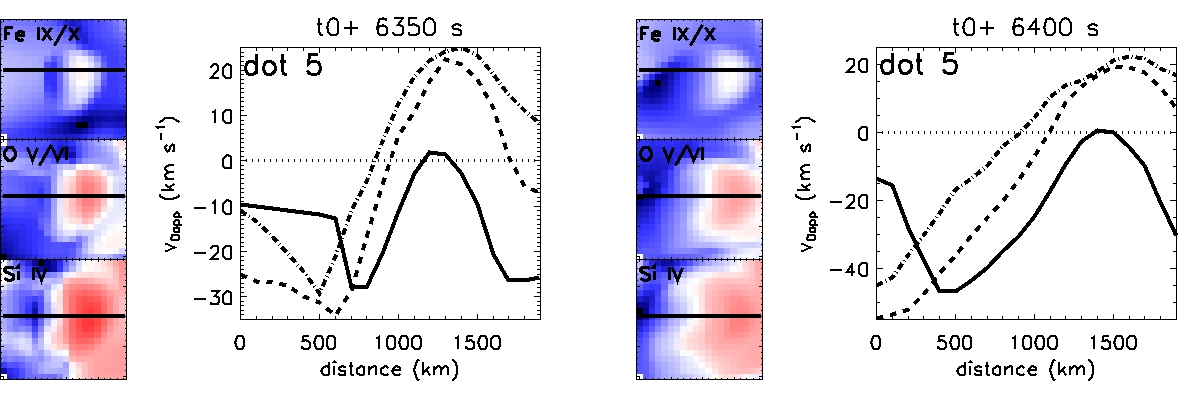}
	\caption{Doppler velocities along a horizontal cut of each of the five example dots shown in Figures \ref{f7_simdots1} and \ref{f8_simdots2}. In the images of the Dopplergrams of dots, zero is gray/white, red is downflow, blue is upflow.  For each example dot two consecutive image frames are plotted to show how the dot evolves in the first 50 s. The FOV for each dot is the same in the Dopplergrams of the three spectral lines. The horizontal solid black line on each Dopplergram image marks the cut along which the Doppler speed is plotted on their right. Solid, dashed, dash-dotted lines in the plots are for \fe, \oxy, and \siiv\ lines, respectively. A dotted horizontal line in each plot marks the zero velocity level. 
	} 
	\label{f10_dopp}
\end{figure*}

For a quantitative picture of Doppler flows of dots in Figures \ref{f7_simdots1} and \ref{f8_simdots2} we take a cut along each dot and plot their Doppler speed along it in Figure \ref{f10_dopp}. To follow the plasma flows, we make the plots of Doppler speeds in two consecutive image frames for each dot. Most dots show redshifts either weak or strong up to 75 \kms\ in \fe\ lines and up to 100 \kms\ or more in \oxy\ and \siiv\ lines. The downflows are always stronger in \oxy\ and \siiv\ lines. 


\subsubsection{Magnetic field distribution and geometry}

\begin{figure*}
	\centering
	\includegraphics[trim=4cm 0cm 11.5cm 0cm,clip,width=0.95\linewidth]{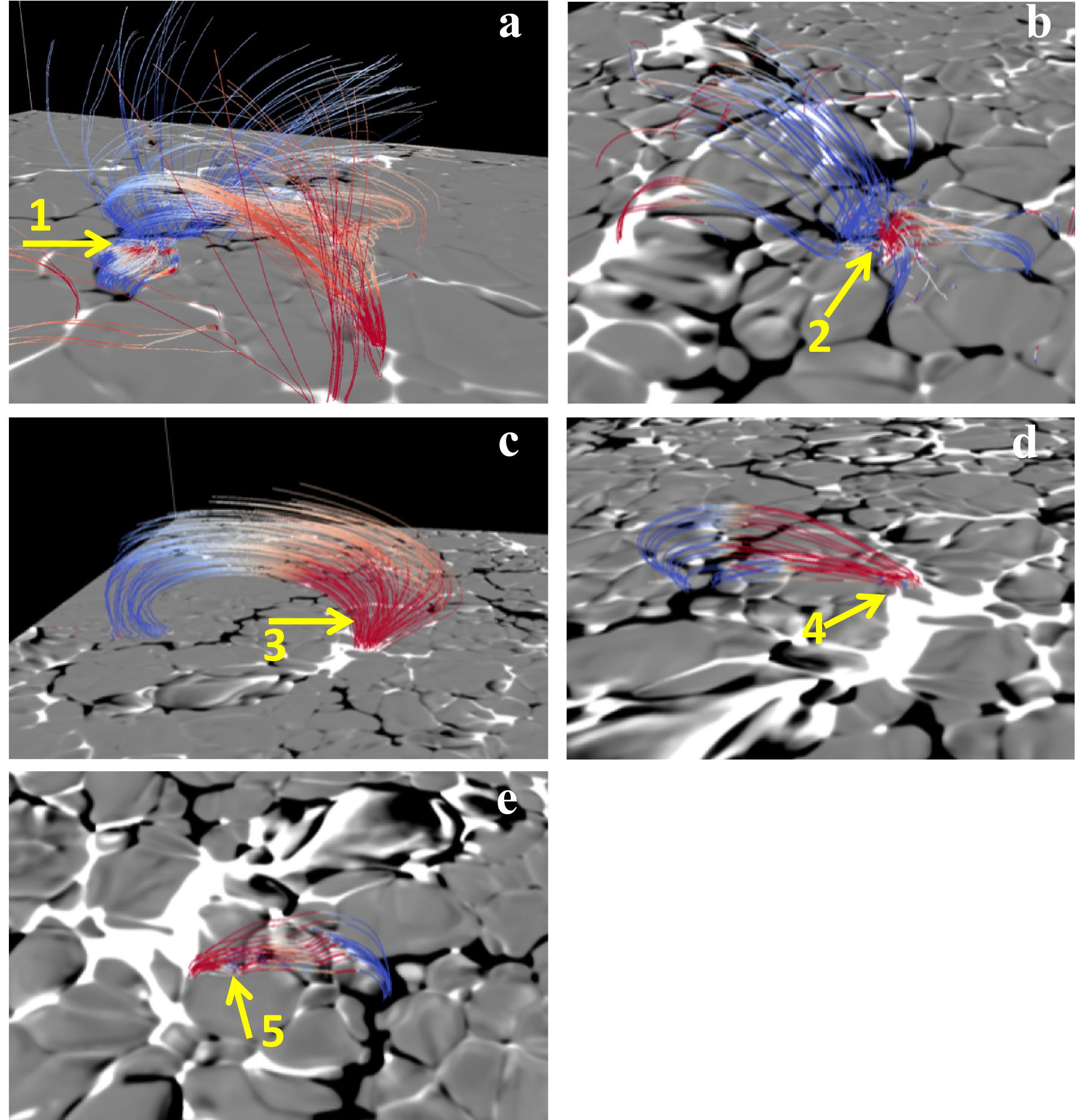}
	\caption{Magnetic field geometry of the five example dots from the simulation. The background images in each panel are Bz maps, saturated at $\pm$500 G. Arrows point to each dot's approximate location. Red and blue colours correspond to positive and negative magnetic fields, respectively. Mixed-polarity magnetic flux is apparently present near four of the five dots (dot 3 is an exception) displaying crossing of the field lines in the lower atmosphere. 
	} 
	\label{f11_field_geom}
\end{figure*}

Consistent with the SDO (AIA+HMI) observations, more than 50\% of the dots that we analyse from the simulation have mixed-polarity magnetic flux at their base, and have sharp neutral lines, or are at the edges of strong magnetic flux patches. This suggests that magnetic reconnection in the lower atmosphere due to the interaction of lower and higher loops is possible. Once magnetic reconnection happens, the resultant lower loops will submerge into the photosphere displaying magnetic flux cancellation in magnetograms. Magnetic reconnection can happen in between the pre-existing and emerging field, or between two existing small loops when the magnetic field gets sheared (as normally visible in Bz maps), creating a suitable magnetic environment for reconnection (e.g., as shown in the Figure 14 of \citealt{tiw19}).  


We traced the magnetic field lines near each of the five example dots to verify if the above idea of dot formation is consistent with their field geometry. For this purpose, we used the visualization software tool VAPOR \citep{li19}. The magnetic field geometry of five example dots is shown in Figure \ref{f11_field_geom}. Red and blue colours in the extrapolated loops correspond to the positive and negative photospheric magnetic field, respectively. Out of our five dots four (dots 1, 2, 4, 5) show field lines interacting closely, at acute angles, low in the atmosphere ($\approx$ 1 Mm from the surface), suitable for magnetic reconnection. These locations of interacting field lines are also the locations where the dots are seated, thus suggesting their formation by magnetic reconnection.  


One of the dots (dot 3) does not show any tangled magnetic field near its location and thus has a possibility of its origin by waves, or downflows. However, this particular dot does not show downflows  in \fe\ lines but does show downflows of 10--20 \kms\ in \oxy\ and \siiv\ lines (Figure \ref{f10_dopp}). Thus, this dot is more likely formed by magneto-acoustic shocks. 
We further analysed this dot in the simulation and found that the dot is indeed formed as a result of magnetic-acoustic waves (or shocks). These waves are generated as a result of nearby flux emergence that perturbs the coronal/transition region plasma. Thus, although photospheric convection pulls the emerging dipole apart, this dot can more directly be linked to the wave motions produced by the interaction of the overlying magnetic field and the emerging magnetic field as it rapidly expands.

\section{Discussion}
We have characterized fine-scale dot-like coronal EUV brightenings observed by SolO/EUI-\hri\ in an emerging magnetic flux region. These dots are tiny entities within a classical CBP. We also analyzed simultaneous SDO/AIA and SDO/HMI data, and compared observed dots with similar bright dots found in a Bifrost MHD simulation of an emerging flux region. During their evolution half of the dots either extend, sometimes explosively, to become a loop, or a surge/jet, or result from and at the end of a loop/surge activity. Some of the brighter and bigger dots may be considered as `dot-like' campfires found in the quiet solar regions \citep{berg21,pane21}.   
Thus, different small-scale coronal dynamic features such as loops, surges/jets, campfires, and any other magnetic structures with plasma flows, most likely exhibit dot-like brightenings at their base, or sometimes on their bulk/apex, during different evolutionary phases. Half of the dots remain isolated during their lifetime and do not show any extensions, and are not accompanied with any of the above structures (e.g., dot `b' in  Figure \ref{f2}).

Somewhat similar (in sizes and lifetimes) EUV bright dots were observed in an active region (unipolar) plage by \cite{regn14} in 193 \AA\ of Hi-C data \citep{koba14}, limited to five minutes of observations. They found their EUV dots to be the foot of much longer coronal loops. 
Most of our dot locations are also at or near the foot of coronal loops but these loops are relatively `short', them being rooted in the initial phase of an emerging bipolar region. Our dots in emerging flux region are likely a result  of magnetic reconnection between the emerging and the pre-existing magnetic field. However, a couple of other possibilities, as discussed below, can not be ruled out. Different dots are consistent with the following three different formation mechanisms. 
 
 \noindent {\bf 1. Magnetic reconnection:} The SDO observations show that most of the bigger and brighter dots (e.g., dots `a' and `c' in  Figure \ref{f4_sdo}) are rooted at strong magnetic field patches, which are often surrounded by opposite-polarity magnetic flux elements, and have sharp polarity inversion lines (PILs). These dots are accompanied with magnetic flux emergence and/or cancellation. Similar to that in observations, in their photospheric magnetograms, many dots in our simulation are either located at a sharp neutral line, or at the edge of a strong magnetic flux patch. Further, many dots in the simulation display both redshifts and blueshifts next to each other (see, e.g., dots 2, 4, and 5), consistent with them being a result of magnetic reconnection. 
 Our simulation shows that the magnetic reconnection happens between the emerging and pre-existing magnetic field in the lower solar atmosphere at $\approx$ 1 Mm above the photosphere. 
 

Magnetic reconnection in the lower solar atmosphere results into a shorter loop and a larger loop \citep{parker79,prie00}. The reconnected-shorter loop submerges into the photosphere if the loop is shorter than a certain length and the magnetic tension dominates over the pressure \citep{van89,moor92,prie14}. As a result of the submergence of this short loop into the photosphere magnetic flux cancellation would be seen \citep[e.g.,][]{tiw14,tiw19}. However, if the resultant loop is long enough so that magnetic tension looses to magnetic pressure then the reconnected loop does not submerge into the photosphere, and no flux cancellation would be seen \citep{prie94,synt20}.

 This magnetic reconnection scenario is well represented in the Figure 14 of \cite{tiw19}, which also demonstrates why many of the dots appear as an extended loop during their evolution. In the case of the longer dashed loop (in Figure 14 of \citealt{tiw19}), sometimes only the reconnection site becomes visible as a dot in the corona -- the extended dashed loop, may remain at much lower, transition region, temperature, as evident in the Bifrost MHD simulation presented here. 
 Thus, dots in \fe\ emission are smaller than in cooler  (\oxy\ and \siiv) lines most likely because only the hottest counterpart of the magnetic reconnection events is visible in the hotter channels. Other parts of the loop reconnection system do not make it to those MK temperatures. The presence of extended structures from dots, during their evolution, further suggests that magnetic reconnection, at the feet of coronal loops (in the chromosphere or TR), is the key cause for generating these dots. 
 
 The magnetic reconnection between emerging and pre-existing magnetic field, resulting into hot EUV plasma blobs and loops, has also been reported recently by \cite{hou21fluxemerg}.   
 Thus, our reconnection idea of dot formation is in general agreement with the scenario of dot formation in \cite{tian14,alp16,tori17} and \cite{tian18}. This reconnection scenario, in some ways, is also consistent with the formation of other small explosive events, including various chromospheric/TR brightenings and surges/jets \citep[e.g.,][]{gupt15,roup17,gosi18,pane18a,pane19,tiw19,shen22}.

  \noindent {\bf 2. Magneto-acoustic waves:} The possibility of some of our dots being generated by magneto-acoustic waves cannot be ruled out. The chromospheric shocks, driven from the photospheric convection, can impact the transition region/lower corona along coronal loops. 
  This scenario is similar to that proposed for bright dots observed  in the transition region by IRIS \citep{mart15,skog16}. It is important to note that EUI 174 \AA\ passband covers \oxy\ lines, which form at a much cooler temperature than 1 MK, and supports the idea of some of the \hri\ dots being at the transition region  temperature and are likely a result of the chromospheric/TR shocks. 
  The line ratios \oxy\ to \fe\ in our simulation (Figure \ref{line_ratios}) show that the \oxy\ lines are equally strong to the \fe\ lines, suggesting that \oxy\ lines could play a significant role in the appearance of the dots observed with \hri, because the wavelength passband is broad and contains all of these lines.
  Furthermore, the footpoints of hot coronal loops in \hri\ 174 \AA\ or AIA 171 \AA\ passbands are often formed in the transition region, not in the corona \citep{delz11}. This possibility is also confirmed by our MHD simulation  in a few cases (e.g., dot 3 in Figure \ref{f11_field_geom}) where no tangled fields suitable for reconnection are found.

\noindent {\bf 3. Impact of downflows:}  Bright dots can also be created by the impact of downflows along coronal loops to the higher density of the chromosphere and transition region. Thus, there will be an increased local density and temperature caused by the impact of those strong downflows on the higher-density lower atmosphere, by shocks or by collision effects. This scenario is similar to that proposed by \cite{klei14,tian14} and \cite{alp16} for some of their dots. The loops in the observations of \cite{klei14} were much longer and the speeds of downflows were supersonic, of the order of 100 \kms. Some of our dots in the Bifrost MHD simulation do show Doppler speeds of close to 100 \kms\ or more, but most of them have a downflow speed of $\le$ 20 \kms, and are not supersonic, particularly in \fe\ emissions.  This could be either due to the limited coronal height of loops in the simulation box, or the fact that both observations and the simulation contain only smaller loops, them being in an initial phase of the emergence. This could probably mean that most, if not all, of the dots in the emerging flux region have likely a different formation mechanism than them being a TR/chromospheric response of downflows. Future, simultaneous IRIS and \hri\ observations of an emerging flux region would help addressing this subject rigorously.

We further note that the above-mentioned studies \citep[i.e.,][]{klei14,tian14,alp16} are focused on the dots in sunspot umbra and penumbra, in much stronger magnetic field regions, as well as in different magnetic topology, than the ones investigated here -- those dots may have a completely different origin. 
The bright coronal dots reported in sunspot penumbra in 193 \AA\ of Hi-C by \cite{alp16} have similar sizes ($\sim$500 km) and horizontal speeds ($<$ 10 \kms), but have much longer lifetimes (270 s) and intensity enhancements (190\% from surroundings). Those dots were proposed to form by magnetic reconnection between the two inclined penumbral magnetic field components (penumbral filaments and spines, see e.g., \#7.1 of \citealt{hino19}) \citep{tian14,alp16}.  The magnetic reconnection scenario proposed for those dots might also work for some of the dots studied here.
 
Although some of our dots (the larger and brighter ones) match with those reported by \cite{tiw19}, in Hi-C  2.1 observations, in that mixed-polarity magnetic flux can be observed at or near the base of dots, Hi-C 2.1 did not show the dots as dim and tiny as observed here. Furthermore, Hi-C 2.1 observations showed much fewer dots than observed in the \hri\ emerging flux region, probably because the active region was at the peak of its lifetime \citep[as discussed in][]{tiw21} and most of the global emergence had already stopped in that active region. 
There might be different reasons for observed differences in the dots in the quiet Sun emerging flux region versus the dots and tiny loops of Hi-C 2.1 in the core of a mature active region. First of all, there simply is not as much reconnection in a mature active region, with ceased flux emergence, as in the emerging flux region in the quiet Sun.
Second, bright surroundings in the active region core might not allow to detect tiny and rather dim dots in the intensity images. Third, about half of the Hi-C 2.1 images were blurred/smeared due to pointing instabilities \citep{rach19}. Fourth, it is quite possible that such tiny events as covered by EUI/\hri\ did not occur during the five minute Hi-C 2.1 observations (or at least during the good image frames). Moreover, the wavelength band of Hi-C 2.1 was broader ($\sim$165--180 \AA) than that of EUI (171--178 \AA), thus possibly capturing more of chromospheric/transition-region emission than that of \hri. A caveat for this argument is that there are not many TR lines between 165 to 170 \AA. 

Previous MHD models have shown that magnetic reconnection between emerging and pre-existing magnetic field can result in the formation of surges/jets \citep{shib92_flux_emergence,yoko95,more13,nobr16}. Because our dots are seen in an emerging flux region, and show extension, the same mechanism might be at work in dots at much smaller scales -- this is what our modelling results, consistently, suggest. As previously mentioned some of the dots are probably the hottest counterparts of jets/surges or loops.

The dots in our study represent the size of the smallest campfires \citep{berg21}. Note that the term campfire represents different coronal brightening events, such as dots, loops, and jets \citep{pane21}. The dot-like campfires have a size of the order of 1000 km and they reside above PILs \citep{pane21}. However, the lifetimes and intensity enhancements of dot-like campfires are much larger than those for our dots. 
Using a triangulation method on simultaneous \hri\ and \sdo/AIA  data, \cite{berg21} and \cite{zhuk21} found that the height of most campfires from the photosphere is $\le$ 5 Mm.  \cite{chen21} proposed, based on their MHD simulation, that component magnetic reconnection generates the largest of campfires, the reconnection taking place at the apex of loops, higher in the corona between 2--5 Mm from the photosphere.
Most dots in our simulation show extension as a loop or jet in \oxy\ and \siiv\ lines, and the brightest part appears as a dot in the \fe\ lines -- this suggests that for dots the reconnection takes place in the lower atmosphere near the TR/chromospheric footpoint of the loop, where flux emergence occurs and the short emerging loop reconnects with the existing (already emerged) loop. The geometrical configuration of dots in our simulation, consistently, show the interaction of short and long loops at a height of $\approx$1 Mm from the photosphere. Thus, some of the properties of campfires are similar to the fine-scale dots investigated here, except that our dots form much lower in the atmosphere, at $\approx$ 1 Mm from the photosphere. 

Thus, our findings also suggest that the heating might not always start from magnetic build-up and triggering at the apex of loops but might often begin at their foot-points, low in the corona/transition region/chromosphere. 

The extension of dots, both in observations and simulation, often appear as a propagation of intensity along a loop, or a small-scale jet at a speed of 30 \kms\ or less. Note that these intensity propagations are still at much smaller spatial and temporal scales than the smallest coronal jets or jetlets reported in the literature \citep{raouafi14,tian14Science,pane18a,pane19,pane20,pane21,chit21,hou21}. 

Because majority of our \hri\ dots in the emerging flux region show coronal (as well as TR) temperatures the presented dots in this study are not Ellerman bombs \citep{elle17,rutt13}. The dots do not show stationary dot-like brightening in AIA 1700 \AA\ either, as noted for EBs \citep{viss19b}. However, some of these could be FAFs \citep{viss15b}. Some of the dots could also be similar to IRIS bombs \citep{pete14}, or UV bursts \citep{youn18,hans17}, but a more extensive investigation is required to settle this issue \citep{hans19}. Again, note that most of our dots have much shorter lifetimes than UV bursts or IRIS bombs ($\sim$5 min), see e.g., \cite{wata11}. Our dots are fine-scale substructures inside a classical CBP, and thus obviously are much dimmer, shorter, and smaller than X-ray/coronal bright points \citep[e.g.,][]{golu74,berg01,madj19}.

It is more likely that dots observed in different UV and EUV wavelengths in different solar environments are generated in many different ways. This discussion is perhaps analogous to the discussion of the nature of solar EUV blinkers \citep{harr97,brko01}. The appearance of IRIS TR images at much higher resolution suggests that when we observe a variety of features driven by very different physical mechanisms we end up with dots, or similar roundish features when seen with the instruments that observe at a much lower spatial resolution.
	
Depending on whether a dot has cooler surrounding (i.e., it is isolated) some dots were disregarded due to not showing up 2-$\sigma$ intensity enhancement that is our selection criterion. That means there might be many more (dimmer) dots than we consider in the EUI images of the emerging flux region. This can be verified, again with future coordinated observations of \hri\ with IRIS.	

Assuming a spherical geometry of dots with an average diameter of 650 km and field strength of 200 G (as found in our simulation), the estimated magnetic energy (B$^2\times$V / 8$\pi$) of dots comes out to be 2.3$\times 10^{26}$ erg. Thus approximate free energy would be in the range of the order of 10$^{26}$ erg (80\% of total magnetic energy), which is on the higher side of that of nanoflares \citep{parker88}. This is similar to the energy estimated for EUV dots in a plage region \citep{regn14}, for nanoflares in small loops \citep{wine13,test13}, and for smaller campfires \citep{pane21}. Thus, our dot-like events have energies capable of heating the corona to million degrees, locally.  CBPs are believed to be major contributors to the quiet corona and these dots mark where exactly the heating happens within CBPs. Further elucidation of fine-scale dots within CBPs in the context of
quiet-Sun coronal heating is obviously of interest.

The EUI/\hri\ has opened a new opportunity to better understand fine-scale coronal explosive events. As SolO gets closer to the Sun better spatial resolution data would be acquired and co-observations with IRIS will be extremely valuable for such investigations as performed here. 
Thus, in future spectral data such as those obtained with IRIS and Hinode/EIS, simultaneous to EUI observations, and high-quality magnetograms such as those obtained with Hinode (SOT/SP), SolO/PHI and DKIST, would provide further insights into the formation of SolO's EUI/\hri\ dots reported here. Of particular interest would be assessing Doppler speeds of dots in different atmospheric heights using IRIS spectra and comparing those with that of Bifrost MHD simulations.  
Furthermore, sophisticated techniques, such as those presented by \cite{hump21} for automatically selecting and characterizing a large number of brightenings, should be used in future on a much larger sample of dots to assess  their common characteristics and corroborate our findings.

\section{Conclusions}
	
Using SolO's EUI/\hri\ 174 \AA\ data, we report on the ubiquitous presence of dot-like fine-scale heating events in and around an emerging flux region. These dots are fine-scale brightening events inside a CBP, and contribute to at least some of their heating. The dots are dim (30\% $\pm$ 10\% brighter than their immediate surroundings), small in size (675$\pm$300 km), short lived (50$\pm$35 seconds), and half of them can be linked with a loop or jet activity of longer span and size. Most of the bigger and brighter EUV dots have a temperature of 1--2 MK, as estimated via DEM analysis of different SDO/AIA passbands, but some are much cooler and might remain at TR/chromospheric temperatures. The line ratios of \oxy\ to \fe\ for dots our simulation suggest that the \oxy\ lines are equally strong to the \fe\ lines. This indicates that \oxy\ lines could play a significant role in the appearance of the dots observed with \hri\ -- as the \hri\ passband is broad, containing all of these lines. 

Many of the \hri\ dots observed in the emerging flux region are probably the hottest counterparts of TR/chromospheric activities, caused by magnetic reconnection. 
The Bifrost MHD simulation of a bipolar flux emergence shows that dots have a bigger extension in TR, cooler, lines, such as \oxy\ and \siiv. Thus, the reconnection site (at $\approx$ 1 Mm from the photosphere) getting hot to MK plasma shows up in \fe\ emission as a dot-like bright transient event. These contain proper motions of  $<$ 10 \kms\ but the intensity propagation along their longer extension, when they extend as a loop or surge/jet, can have a speed of up to 30 \kms. Dots in the simulation often contain mixed Doppler signals in \fe\ emission, both blueshifts and redshifts of the order of 10 \kms, but Doppler speeds can be multiple times larger. Redshifts are always stronger in \oxy\ and \siiv\ lines than in \fe\ lines. The magnetic field geometry of dots in our simulation suggests that most dots are caused by magnetic reconnection between emerging and pre-existing magnetic field -- thus also suggesting that heating in a loop does not always start at the loop's apex, rather can often start near their TR/chromospheric feet. Thus, our observational and modelling results suggest that magnetic reconnection in these dots plays an important role in some of the coronal heating of emerging flux regions and provide new insights into the heating at fine-scales by magnetic reconnection. 

Because magnetic reconnection happens low in the TR/chromosphere, the presence of mixed-polarity magnetic flux, and flux cancellation as a result of the submergence of the lower reconnected loops, is consistent with the findings of \cite{tiw19} for dot-like, loop-like and surge-like events in the core of the Hi-C 2.1 active region, and of \cite{pane21} for dot-like, loop-like, complex, and jet-like campfires in the quiet solar corona. Some dots could well be caused by chromospheric shocks, either directly driven from the photospheric convection, or generated from the interaction of emerging and overlying field. A small percentage of dots could also be a response of the impact of downflows along coronal loops on to the TR/chromospheric density. For this, a further detailed investigation is required.


 The spatio-temporal filling factor of these dots has yet to be determined.
Further, dots found in different magnetic environments and regions in the solar atmosphere may have different formation mechanisms. Whether EUV dots are limited to strong field loops such as those found in plage areas \citep{regn14}, sunspots \citep{tian14,klei14,alp16,deng16}, core of ARs \citep{tiw19}, and emerging flux regions (this work), or are present at the base of each coronal loop even in the weaker magnetic regions such as in the quiet Sun, coronal holes, as well as in other solar features such as filaments, and plumes, remains to be seen.


	\vspace{0.5cm}
We would like to thank the referee for carefully reading our manuscript and for constructive suggestions. We thank Mark Cheung (LMSAL) for discussions about DEM analysis. S.K.T. gratefully acknowledges support by NASA HGI award (80NSSC21K0520) and NASA contract NNM07AA01C (Hinode). 
V.H.H. is supported by NASA grant 80NSSC20K1272: Flux emergence and the structure, dynamics, and energetics of the solar atmosphere. 
B.D.P. was supported by NASA contract NNG09FA40C (IRIS).
N.K.P's research was supported by NASA grant NNG04EA00C (SDO/AIA) and HGI award (80NSSC20K0720). 
Solar Orbiter is a space mission of international collaboration between ESA and NASA, operated by ESA. The EUI instrument was built by CSL, IAS, MPS, MSSL/UCL, PMOD/WRC, ROB, LCF/IO with funding from the Belgian Federal Science Policy Office (BELSPO/PRODEX PEA 4000112292); the Centre National d'Etudes Spatiales (CNES); the UK Space Agency (UKSA); the Bundesministerium f\"ur Wirtschaft und Energie (BMWi) through the Deutsches Zentrum f\"ur Luft- und Raumfahrt (DLR); and the Swiss Space Office (SSO). IRIS is a NASA small explorer mission developed and operated by LMSAL with mission operations executed at NASA Ames Research Center and major contributions to downlink communications funded by ESA and the Norwegian Space Centre. 
The AIA and HMI data are courtesy of NASA/SDO and the AIA and HMI science teams. We acknowledge imagery produced by VAPOR (\url{www.vapor.ucar.edu}), a product of the Computational Information Systems Laboratory at the National Center for Atmospheric Research. 
This research has made use of NASA's Astrophysics Data System and of IDL SolarSoft package.


\newpage

\appendix
\section{Further examples of dots in the HRI$_{\rm EUV}$ observations of emerging flux region}\label{more_hri_dots}
\begin{figure*}[hp]
	\centering	
	\includegraphics[trim=3.95cm 4.1cm 3.3cm 1.6cm,clip,width=0.88\linewidth]{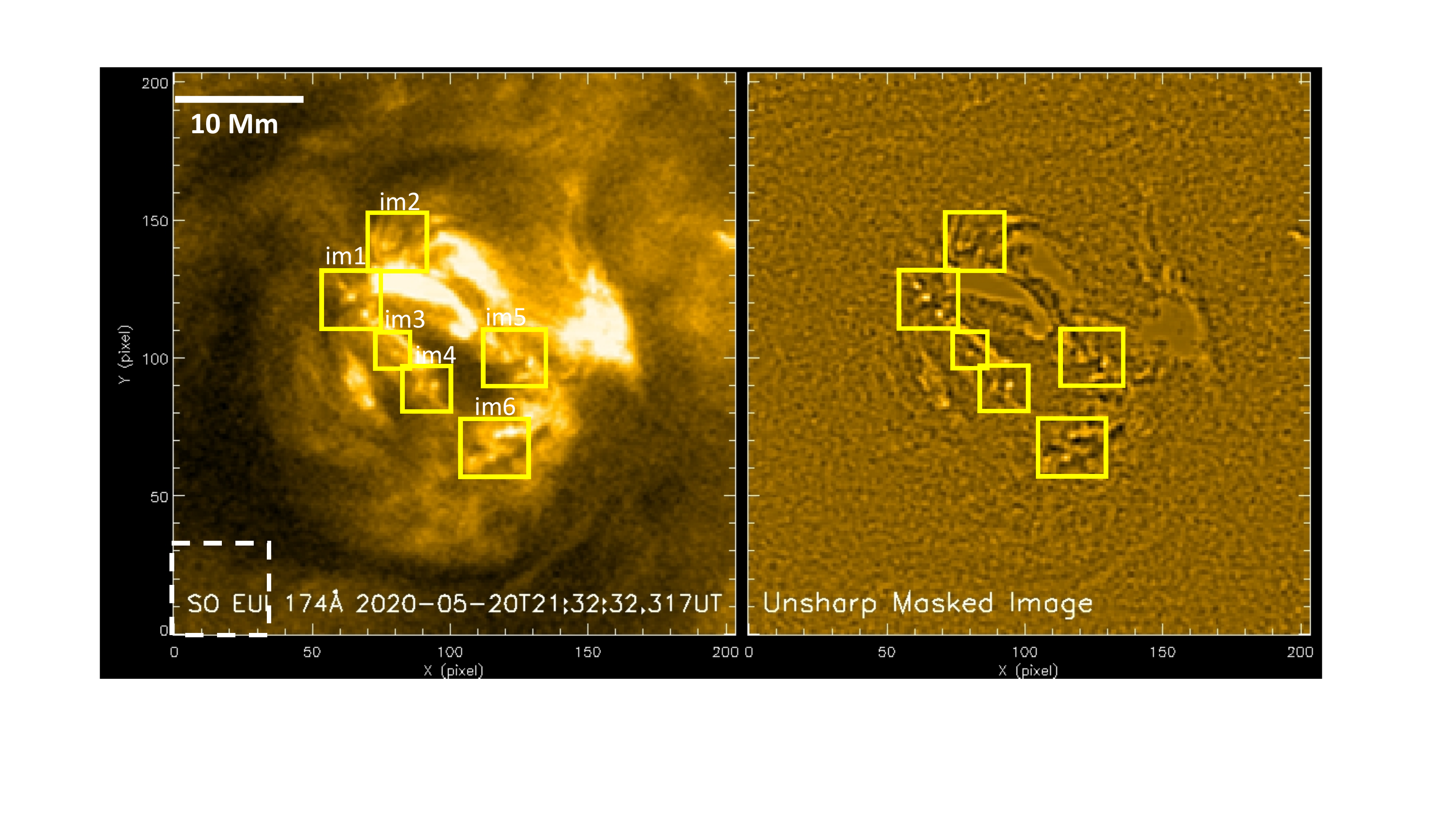} 
	\includegraphics[trim=3.15cm 2.75cm 2.25cm 2cm,clip,width=0.88\linewidth]{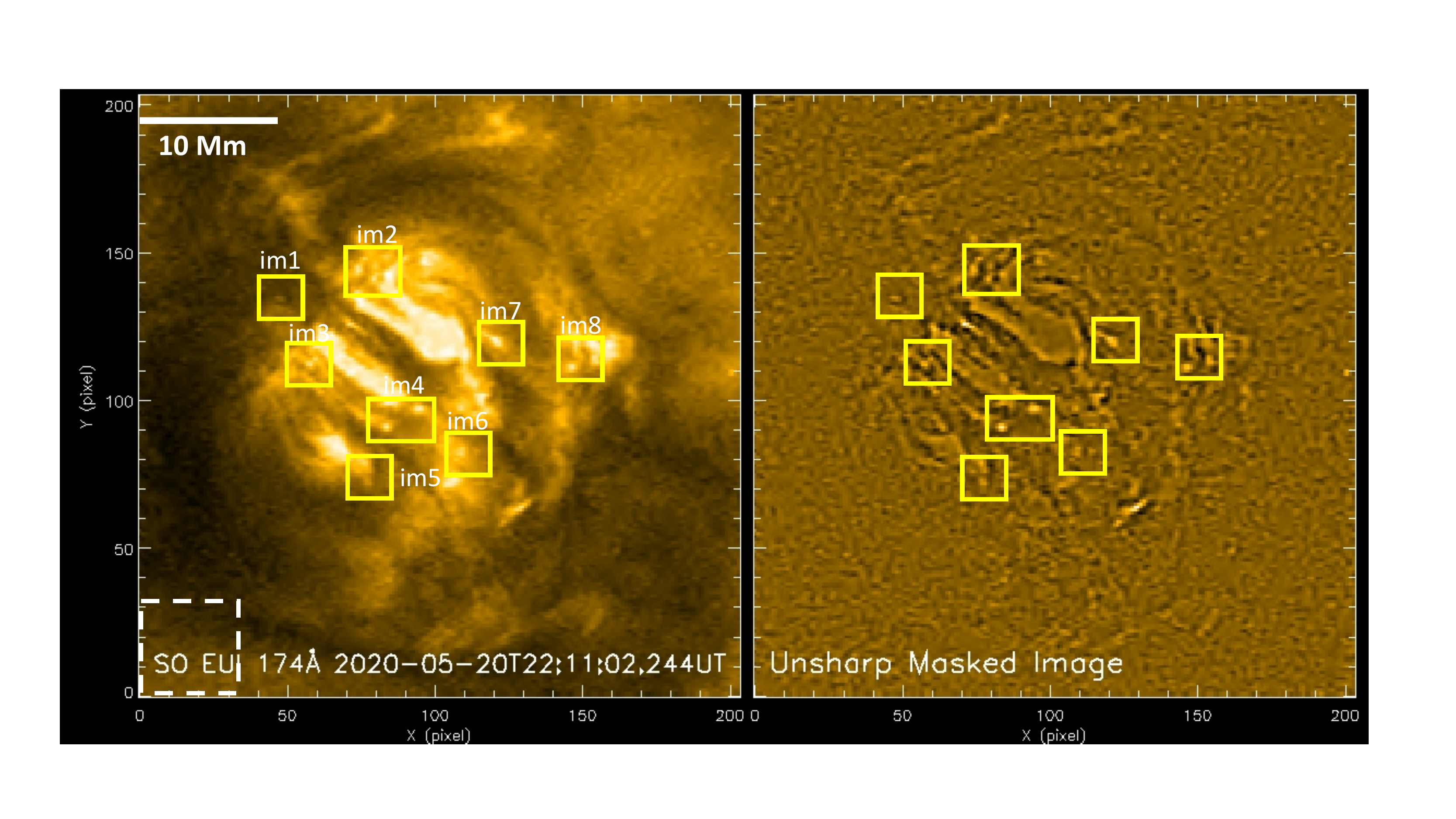}
	\caption{Additional examples of dots in EUI/HRI$_{\rm EUV}$ observations. The left panel in each row is HRI$_{\rm EUV}$ 174 \AA\ image, and the right panel is unsharp masked image of it. Different boxes outline the regions of selected dots in each image frame. A white dashed box on the bottom left of the left panel outlines the region that is used for noise estimation. A white horizontal bar on the 174 \AA\ image scales 10 Mm distance, for reference.}
	\label{f12}
\end{figure*}

In the following, Figure \ref{f12},  we show two additional image frames from the ``movie1.mp4" with many dots outlined by yellow boxes. 


\newpage

\section{SDO/AIA images, corresponding to Figure 4, in different AIA channels}\label{iris_dots}

Here we show images in different AIA channels of the same time and FOV as shown in Figure \ref{f4_sdo} for 171 \AA. The three dots show faint signatures in the AIA 304, 193, and 131 \AA\ images, but are not evident in AIA 1600, 1700, and 94 \AA\ images. 

\begin{figure*}[h]
	\centering
		\includegraphics[trim=0cm 0cm 0cm 0cm,clip,width=\textwidth]{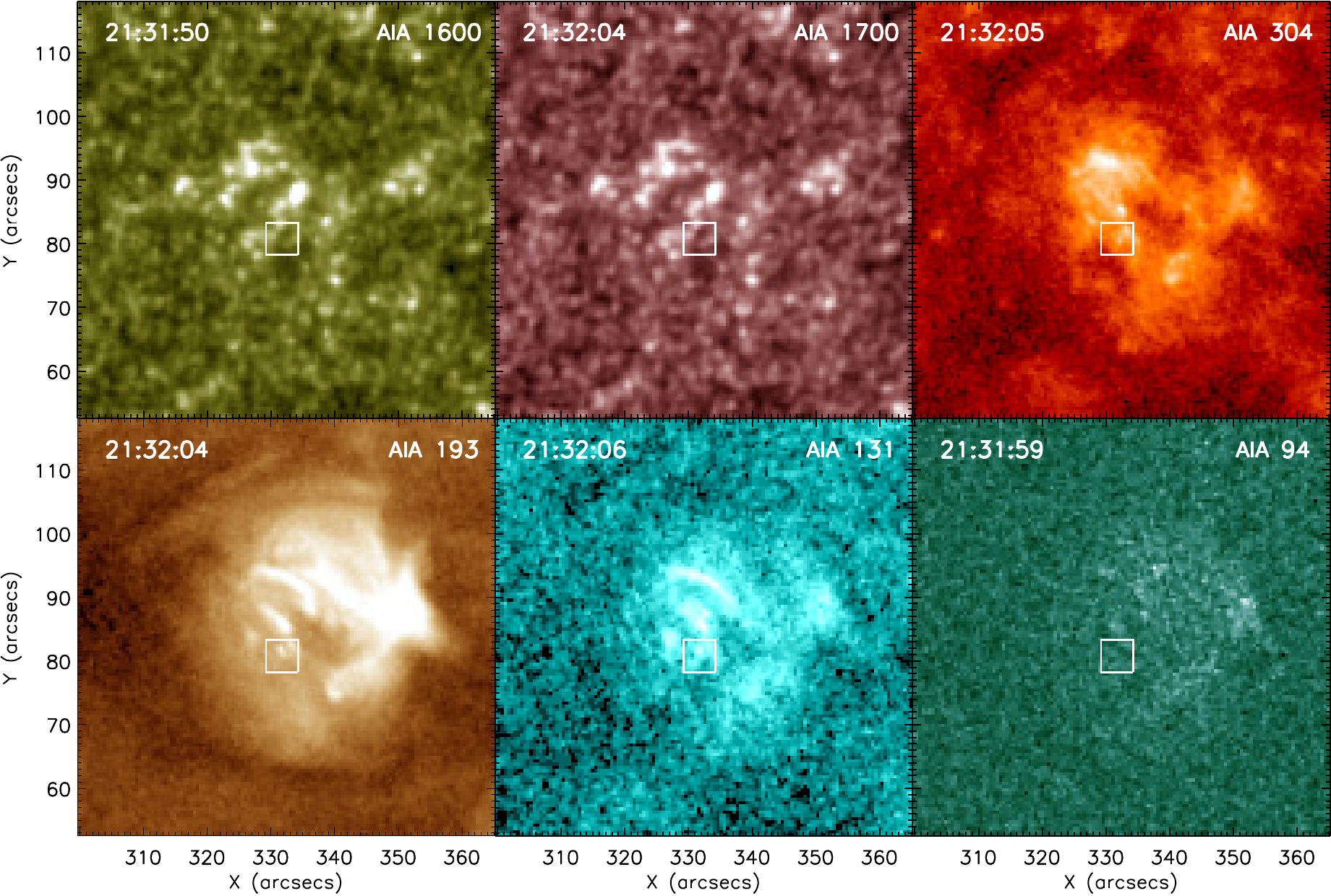}
	\caption{Images corresponding to Figure \ref{f4_sdo}, in different AIA channels. From left to right, AIA 1600, 1700, 304 \AA\ in the upper row, and AIA 193, 131, and 94 \AA\ in the lower row. A white box in each image is the same as that in Figure \ref{f4_sdo}.  Note that, similar to those in Figure \ref{f4_sdo}, these AIA images are de-rotated to the central image time, that is at 20-May-2020 21:44:51 UT. A roll angle correction of 6\degree\ is made to match that with \hri.}
	\label{aia_channel}
\end{figure*}

\newpage
\section{Dots in IRIS observations of an emerging flux region}\label{iris_dots}
We note that there was no IRIS co-observations with the EUI/HRI$_{\rm EUV}$ data used in the present work. We looked for independent IRIS observations capturing initial phases of magnetic flux emergence, to see if there are  fine-scale dots in these observations of TR/chromospheric lines. 
 
Here, we show an example map from IRIS Si IV 1400 \AA\ SJI observations of an emerging flux region, with some dots outlined inside a few boxes on it. The unsharp masked image and corresponding \sdo/HMI LOS magnetograms are also displayed.

\begin{figure*}[h]
	\centering
	\includegraphics[trim=0.1cm 6.3cm 0.1cm 1.4cm,clip,width=\textwidth]{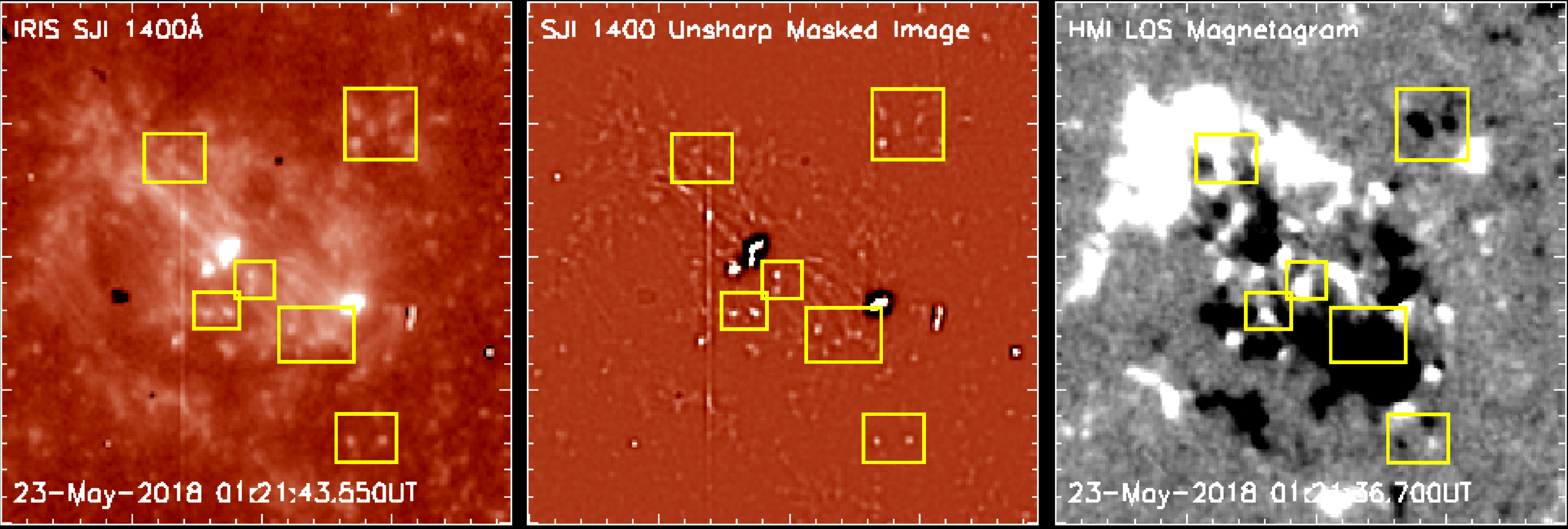}
	\caption{IRIS SJI 1400 \AA\ image (left panel) and its unsharp masked image (middle panel) of an emerging flux region, displayed together with \sdo/HMI LOS magnetogram (right panel) that is the closest in time to the SJ 1400 image. Many small-scale dots can be noticed, some outlined by yellow boxes for easy identification. The presence of dots in \siiv\ lines similar to that of EUI suggests that they could be formed in the transition region. However a detailed study comparing dots one-to-one in the transition region and corona will be required to confirm this.
	}
 \label{iris_dots}
\end{figure*}

\newpage 
\section{Bifrost MHD simulation: synthetic \oxy\ and \siiv\ lines}\label{sim_osi}
Here, we plot \oxy\ 172/173 \AA\ and \siiv\ 1393 \AA\ images of the corresponding \fe\ image frame shown in Figure \ref{f6_sim}. 

\begin{figure*}[h]
	\centering
	\includegraphics[trim=0cm 0cm 0cm 0cm,clip,width=\textwidth]{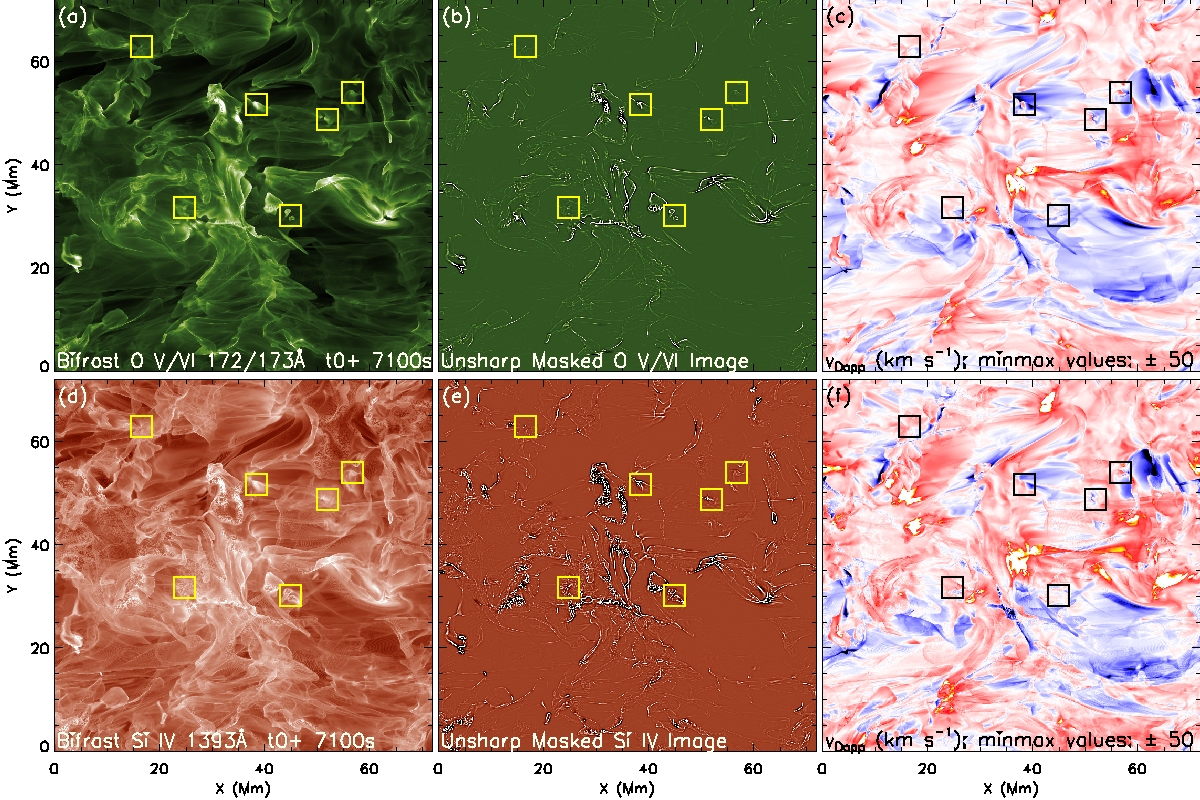}
	\caption{Example image frame (same as in Figure \ref{f6_sim} for \fe\ lines) of \oxy\ and \siiv\ lines. The  dot locations outlined by the yellow boxes are the same as in Figure \ref{f6_sim}. 
	}
	\label{osi_context}
\end{figure*}

\end{document}